\documentclass[11pt]{article}

\usepackage{putex}

\RequirePackage[a4paper,top=30.6mm,bottom=38.6mm,left=34.6mm,right=34.6mm,footskip=1.3cm]{geometry}
\usepackage{setspace}
\usepackage{booktabs}
\usepackage{caption}
\usepackage{amsmath}
\usepackage{enumerate}
\usepackage{cite}
\usepackage{slashed}
\usepackage[utf8]{inputenc}
\usepackage{rotating}
\usepackage[
colorlinks=true,
linkcolor=black,
urlcolor=blue,
filecolor=black,
citecolor=red,
]{hyperref}

\numberwithin{equation}{section}

\def \< {\left<}
\def \> {\right>}

\newcommand{\be}{\begin{equation}} \newcommand{\ee}{\end{equation}}
\newcommand{\bea}{\begin{eqnarray}}  \newcommand{\eea}{\end{eqnarray}}
\newcommand{\nn}{\nonumber}

\usepackage{amsfonts, amsthm}
\usepackage[english]{babel}
\usepackage{slashed}
\usepackage{mathrsfs}
\usepackage{amssymb}
\usepackage{color}

\newcommand{\ma}[1]{\mbox{$\mathcal{#1}$}}

\newcommand{\mrm}[1]{\mbox{$\mathrm{#1}$}}

\begin{document}

	\begin{center}        
		\Huge Metrics over multi-parameter \\ AdS vacua 
	\end{center}
	
	\vspace{0.7cm}
	\begin{center}        
		{\Large  Eran Palti$\;^1$\quad and\quad Nicol\`o Petri$\;^2$}
	\end{center}
	
	\vspace{0.15cm}
	\begin{center}  
		\emph{$^1$ Department of Physics, Ben-Gurion University of the Negev,}\\
		 \emph{Be'er-Sheva 84105, Israel}\\[.3cm]
		\emph{$^2$ INFN sezione di Torino, via Pietro Giuria 1, 10125 Torino, Italy}\\[.3cm]
		\emph{}\\[.2cm]
		e-mails:  \tt palti@bgu.ac.il, \; nicolo.petri@to.infn.it
	\end{center}
	
	\vspace{1cm}
	
	
	\begin{abstract}
	\noindent
	We study the notion of a distance between different AdS vacua of string theory. The distance is measured by a metric that is derived by taking an off-shell quadratic variation of the effective action, and evaluating it over families of vacua. We calculate this metric for increasingly complex families of vacua. We first consider the two-parameter families of solutions of type $\mathrm{AdS}_4 \times \mathbb{C}\mathrm{P}^3$. We find that the metric is flat and positive, and so yields a well-defined distance along any path in the space of solutions. We then consider solutions of type $\mathrm{AdS}_3 \times S^3 \times \mathrm{CY}_2$ which have two (non-compact) flux parameters as well as a moduli space. We find that the space of solutions factorises between directions which vary the AdS radius, and the moduli space. The metric over AdS variations is flat and positive, and the metric over the moduli space is the usual one. Finally, we consider solutions of type $\mathrm{AdS}_3 \times S^3 \times S^3 \times S^1$ which also have a further direction in the space of solutions that is compact. We find that the metric is flat only on non-compact directions in the space of solutions. Restricting to such directions, we evaluate the metric and find it is positive definite and therefore yields a well-defined distance along any path. 
	\end{abstract}
	
	\thispagestyle{empty}
	\clearpage
	
	\tableofcontents
	
	\setcounter{page}{1}

	\section{Introduction}
	\label{sec:intro}

Vacua of string theory often come in families. One can consider two distinctive types of such families. The first type corresponds to vacua which are related by varying the expectation values of fields with no potential, so by moving along a moduli space. There is a notion of a metric, and an associated distance, over such vacua. The distance between vacua in moduli space is known to have an important role in the Swampland program, see \cite{Ooguri:2006in,Baume:2016psm,Klaewer:2016kiy,Grimm:2018ohb,Grimm:2018cpv,Corvilain:2018lgw,Blumenhagen:2018nts,Lee:2018spm,Lee:2019wij} for some of the early work on the topic, and \cite{Palti:2019pca,vanBeest:2021lhn} for reviews. The distance along moduli space is well-defined and understood for both Minkowski and AdS vacua of string theory. In the latter case, it is holographically dual to a distance along the conformal manifold of a CFT. 

The second type of families of vacua are unique to AdS, and correspond to varying cosmological constants, or AdS radii. While it is natural to expect that a good notion of distance should exist between such vacua, it remains an open problem to find the correct prescription for it. Also holographically such a distance is not well understood, since it would correspond to a notion of distance between different CFTs (for example, with different central charges), rather than exactly marginal deformations of the same CFT. 

In \cite{Li:2023gtt}, following initial ideas in \cite{Lust:2019zwm}, a certain prescription for calculating distances between AdS vacua with different cosmological constants was proposed.\footnote{See \cite{Basile:2023rvm,Mohseni:2023ogd,Farakos:2023wps,Shiu:2023bay,Mohseni:2024njl,Demulder:2024glx,Debusschere:2024rmi} for alternative ideas on how to define such a notion of distance.} The idea was to take a quadratic variation of the action, and evaluate it on the family of solutions. The coefficient of the quadratic variation is then taken as the metric on the space of solutions. This is actually the same as the prescription for calculating the moduli space metric, but it involves some subtleties that were addressed in \cite{Li:2023gtt}. One particularly interesting property of the resulting metric is its sign. A variation of solely the AdS radius yields a negative metric, which is a manifestation of the famous conformal factor problem in Euclidean quantum gravity. However, it was found in \cite{Li:2023gtt,Palti:2024voy}, that the variations of the other parameters in the solution, as required to stay within the space of solutions, make the overall metric positive and so yield a well-defined distance. Indeed, it was conjectured in \cite{Palti:2024voy} that this positivity of the metric is a property of any family of solutions of quantum gravity, so is a Swampland constraint. 

In this work we continue the study of metrics and distances on AdS vacua by applying the general prescription of \cite{Li:2023gtt} to increasingly complex setups. Our main interest is in multi-parameter families of solutions. This means that the space of solutions is not one-dimensional, and so there are different paths that one can take in that space. This setup therefore provides a check on whether the positivity of the metric depends on the path. We find that the metric over the full space of solutions, so evaluated along any path, is flat and positive definite. More precisely, it is flat over the space of solutions which come in infinite families. We show this for three types of families of solutions, based on the geometries $\mathrm{AdS}_4 \times \mathbb{C}\mathrm{P}^3$, $\mathrm{AdS}_3 \times S^3 \times \mathrm{CY}_2$ and $\mathrm{AdS}_3 \times S^3 \times S^3 \times S^1$. The latter geometry also has a parameter which controls solutions that come only in a finite family (within the supergravity regime), so parameterises a compact direction in the space of solutions. The metric over this compact parameter is not flat.

 An interesting property of the solutions on $\mathrm{AdS}_3 \times S^3 \times \mathrm{CY}_2$ and $\mathrm{AdS}_3 \times S^3 \times S^3 \times S^1$ is that they have a moduli space. A component of the full metric on the space of solutions is then the metric on the moduli space. This component is calculated in the same way as the other metric components, and agrees with the known result. We find that the full metric factorizes over the modulus and non-modulus directions in solution space. The metric over the modulus direction is the standard one. The prescription for calculating the metric over non-modulus directions needs to be slightly modified in the presence of moduli. More precisely, there is a certain new ambiguity in the procedure in the presence of moduli, associated to the ordering of the variations and the Weyl rescalings, which needs to be fixed in a specific way. 
  
 The paper is structured as follows. In section \ref{sec:metricoverads} we review, and slightly extend, the general procedure for calculating metrics over AdS vacua introduced in \cite{Li:2023gtt}. In section \ref{sec:ads4cp3} we study $\mathrm{AdS}_4\times \mathbb{CP}^3$ families of solutions. These are multi-parameter backgrounds, with purely non-compact directions (infinite families of solutions), and no moduli. In section \ref{sec:weylmoduli} we discuss the issue of moduli in a general way, developing the prescription for calculating the AdS distances when moduli are present. In section \ref{AdS3vacua} we study $\mathrm{AdS}_3\times S^3\times \mathrm{CY}_2$ families of solutions. These are multi-parameter backgrounds which also have a modulus direction. In section \ref{sec:ads3s3s3s1} we study $\mrm{AdS}_3\times S^3\times S^3\times S^1$ families of vacua. These have multi-parameter non-compact directions in the solutions space, and also a modulus direction, and further a compact direction in the solutions space, so a finite family. We summarise our results in section \ref{sec:dis}. 

\section{The metric over AdS string vacua}
\label{sec:metricoverads}

In this section we summarize the prescription developed in \cite{Li:2023gtt,Palti:2024voy} for defining a consistent notion of metric and distance over AdS string vacua. It is primarily a review of those works, but also includes some new general results which are relatively simple extensions. 

We consider a family of solutions that are compactifications of string theory to $d$-dimensional AdS$_d$ vacua. The procedure for calculating a metric over the family of solutions is to give an infinitesimal spacetime dependence to the parameters in the solutions which vary within the family. We then calculate a $d$-dimensional action for those perturbed parameters, and the metric is the coefficient in this action in front of the quadratic variation terms. 

More precisely, it is the coefficient in front of the kinetic term resulting from identifying the different variations according to how they are related within the family of solutions. For example, we can consider varying the AdS radius, and also the internal dimensions radii, and then calculate kinetic terms for them both. But within a family of solutions they would be related, and so after imposing this relation we would obtain only a single kinetic term. If they are the only parameters which vary within the family, the metric is the coefficient in front of that term. If there are more parameters, then we calculate their effective kinetic terms, and then identify them again using their relations to the other parameters in the family of solutions. This means that if the family of solutions is a one-parameter family, then there is a single kinetic term in the action and the metric is its coefficient. If the family of solutions involves multiple independent parameters, then we would have multiple kinetic terms, and the metric would have matrix components. So a family of solutions with $N$ independent parameters has an $N$-dimensional space of solutions, and an $N$-dimensional metric on that space. In \cite{Li:2023gtt,Palti:2024voy}, only one-dimensional families of solutions were considered. In this work we consider two, three, and four-dimensional families of solutions. 

Note that we do not assume separation of scales between the AdS radius and the internal dimensions radii, so this is not an effective low-energy action. Rather, the action is just a way to encapsulate information on the full higher-dimensional solution. That information is the equations of motion for variations of specific parameters of the solution. The action is the one associated to those parameters. There is no sense in which those parameters are lighter than other possible variations of the solution away from the specific family of solutions we are considering, rather we simply turn off all other variations by hand. 

Following \cite{Li:2023gtt,Palti:2024voy}, we denote the procedure of imposing the relations between the different parameters, implied by the equations of motion of a given family of solutions, as going {\itshape on-shell}. When we vary the different parameters independently, we say that we are working {\itshape off-shell} (or that we {\itshape gauge} a parameter of the solution). 
In what follows we discuss such off-shell variations for some universal parameters.

\subsection{AdS radius variations}
\label{AdSvolumevariations}

Consider first the variation of the AdS radius. In a given family of solutions, the AdS radius varies with the vacuum energy $\Lambda$ associated to the various sources. We can write this as a conformal variation of the metric, with parameter $\sigma$, as
\begin{equation}\label{AdSvolumeVariations}
 ds_d^2=e^{2\sigma}\,d\hat s^2_d\qquad \text{with} \qquad \Lambda=e^{-2\sigma}\,\hat\Lambda\,,
\end{equation}
where $ds^2_d$ and $d\hat{s}^2_d$ are the metrics of two different AdS spaces within the family of solutions, with associated vacuum energies $\Lambda$ and $\hat\Lambda$. 
To extract a metric over such variations in the solutions, we let $\sigma$ have some (infinitesimal) dependence on the AdS spacetime, and then extract a kinetic term from the gravity action. Schematically we can write that
\begin{equation}
 \text{on-shell:}\qquad \sigma=\text{const}\qquad \longrightarrow \qquad \text{off-shell:}\qquad \sigma=\sigma (x)\,,
\end{equation}
where $x$ represents a generic spacetime direction(s) within the AdS vacuum.
Such a procedure implies that the equations of motion are no longer satisfied. In fact the AdS geometry gets infinitesimally deformed when $\sigma$ is promoted to a local function.

We can now evaluate the gravity action over the variations \eqref{AdSvolumeVariations} and extract the metric over vacuum energy variations. We point out that once the metric is computed, we can go back on-shell taking constant $\sigma$. To this aim one can start from the Einstein-Hilbert action and derive the following result\footnote{We use Planck units. So $M_p=1$, with $M_p$ the Planck mass in lower dimension $d$.} \cite{Li:2023gtt,Palti:2024voy}
\begin{equation}
\begin{split}\label{offshellActionExternalVariations}
 &S_d =\frac12\int d^dx\sqrt{-g_d}\,R_d=\frac12\int d^dx\sqrt{-g_d}\,\left(\tilde R_d-K_{\sigma\sigma}(\partial \,\sigma)^2 \right)\,,\\
 &K_{\sigma\sigma}=-(d-1)(d-2)\,,
 \end{split}
\end{equation}  
where we introduced the notation $\tilde R_d=e^{-2\sigma}\hat R_d$ and we dropped total derivatives.
We stress that the above action is defined over off-shell configurations.\footnote{The action \eqref{offshellActionExternalVariations} is obtained by using (\ref{weylformula}), integrating by parts, and absorbing the exponential factors in $\sigma$ within the kinetic term and the volume factor using the relations $e^{-2\sigma}\,\hat g_d^{mn}=g_d^{mn}$ and $e^{d\sigma}\,\sqrt{-\hat g_d}=\sqrt{- g_d}$. See \cite{Li:2023gtt} for more details. We use the compact notation $(\partial \sigma)^2=g_d^{mn}\partial_m\sigma\partial_n\sigma$.} Once we go back on-shell, evaluating $\sigma$ to a constant, we obtain the standard gravity action.

The coefficient $K_{\sigma\sigma}$ in \eqref{offshellActionExternalVariations} is the metric element over the volume variations of the AdS space. We point out that $K_{\sigma\sigma}$ is negative. This result is related to the negative contribution to the Einstein-Hilbert action from the conformal mode of the metric, which is the cause of the famous conformal factor problem of Euclidean quantum gravity \cite{Gibbons:1976ue,Marolf:2022ybi}. 

It is important for our analysis to stress that Weyl rescalings are not the unique transformations that can realize AdS volume variations. To show this it is helpful to introduce the Poincar\'e coordinates for the unit radius AdS geometry,
\begin{equation}
\begin{split}\label{AdSPoincarecord}
 &d\hat s^2_{d}=z^{-2}\bigl(ds^2_{M_p}+dz^2\bigr)\,\qquad \text{with}\qquad\text{vol}_{d}=z^{-d}\,dx^0\wedge \cdots \wedge dx^{p}\wedge dz\,,
 \end{split}
\end{equation}
where $ds^2_{M_p}=-(dx^0)^2+\dots+(dx^{p})^2$ is the metric on the $p$-dimensional Minkowski slicing $M_p$ and $\text{vol}_{d}$ is the volume of (unit radius) AdS. Consider the following class of deformations, depending on two parameters $\sigma$ and $\sigma_1$, of the AdS geometry \cite{Li:2023gtt}
\begin{equation}\label{AdSvariationssigma1}
\begin{split}
 &ds_{d}^2=e^{2\sigma} d\hat s^2_{d}\,,\qquad \text{with} \qquad d\hat s^2_d=\frac{1}{z^2}\left(ds^2_{M_p}+e^{2\sigma_1}dz^2\right) \,.
 \end{split}
\end{equation}
As extensively discussed in \cite{Li:2023gtt,Palti:2024voy}, this type of variation is particularly important for a consistent metric over the variations of AdS vacua.
First we note that, when $\sigma$ and $\sigma_1$ are evaluated to constant values, one recovers the AdS metric in Poincar\'e coordinates \eqref{AdSPoincarecord}. This can be observed by reparametrizing the radial coordinate as $\bar z=e^{\sigma_1}\,z$. This leads to the AdS metric \eqref{AdSPoincarecord} with the radius shifted as $ds_{d}^2=e^{2\sigma+2\sigma_1} d\hat s^2_{d}$. On the other hand, when $\sigma$ and $\sigma_1$ are allowed to vary locally (so are gauged), the above deformations are not equivalent to variations with $\sigma_1=0$ considered previously.

We can now compute the action for the variations \eqref{AdSvariationssigma1}. To do so we restrict our focus to variations which depend only on the $z$ coordinate in AdS, so take $\sigma=\sigma(z)$ and $\sigma_1=\sigma_1(z)$. These conditions will be necessary in order to reproduce a well-defined metric, once flux variations are included in the analysis. The action is then given by \cite{Li:2023gtt}\footnote{For the derivation of the action \eqref{offshellactionMetricSigma1} we refer to Appendix C in \cite{Li:2023gtt}.}
\begin{equation}
\begin{split}\label{offshellactionMetricSigma1}
 S_{d}=\frac12\int d^dx\sqrt{-g_d}\,R_d=\frac12\int d^dx\,\,\sqrt{- g_{d}}&\bigl(\tilde R_{d}-2pe^{-2(\sigma+\sigma_1)}z\,\partial_z\sigma_1-K_{\sigma\sigma}(\partial \,\sigma)^2\bigr)\,,
 \end{split}
\end{equation}
where $\tilde R_d$ is defined as the Ricci scalar of the AdS space with radius $e^{\sigma+\sigma_1}$. The metric element $K_{\sigma\sigma}$ is the same as in \eqref{offshellActionExternalVariations}. Therefore, the inclusion of $\sigma_1$ only produces a new contribution that is linear in $\partial_z\sigma_1$. As we will see, this term plays a crucial role in the formulation of a consistent metric including flux variations.

\subsection{Internal volume and dilaton variations}
\label{Sec:internalvolume}

Further universal features of families of AdS vacua in string theory are the variations of the radius of the internal compact dimensions, and of the dilaton (string coupling). 
Concretely, we can consider the variation of the internal volume in higher-dimensional geometries of the type
\begin{equation}
 \text{AdS}_d\times Y_{	k}\,,
\end{equation}
where $Y_k$ is a compact manifold. We can consider the following family of metrics,
\begin{equation}
 ds_{d+k}^2=e^{2\sigma}\,d \hat s^2_d+e^{2\tau}\,d\hat s^2_k\,,
\end{equation}
where $d \hat s^2_d$ and $d \hat s^2_k$ are respectively the metrics over (unit radius) AdS$_d$ and over $Y_k$. We want now to study the variations of the parameters $\sigma$ and $\tau$.
Since we are interested in string theory vacua, we also include in the analysis a $(d+k)$-dimensional dilaton $\Phi$. 

Following the approach of previous section, we can now suppose that $\sigma$, $\tau$ and $\Phi$ take an infinitesimal profile on the AdS background and compute the action for such variations. Let us consider the {\itshape string frame} action
\begin{equation}
 \begin{split}\label{EHdilatonAction}
  S=\frac{1}{2\kappa_{d+k}^2}\int d^{d+k}x\sqrt{-g}\,e^{-2\Phi}\left(R+ 4g_d^{mn} \partial_m \Phi\, \partial_n\Phi \right)\,,
 \end{split}
\end{equation}
where $\kappa^2_{d+k}$ is the higher-dimensional gravitational coupling. Since we are looking at the metric variations in the lower-dimensional theory, we need to reduce the $(d+k)$-dimensional action to $d$-dimensions, fixing the higher-dimensional gravitational coupling as $\kappa^2_{d+k}=\text{Vol}(Y_k)=\int d^k y \sqrt{\hat g_k}$ with $\{y\}$ coordinates over $Y_k$. This result is derived in \cite{Palti:2024voy}. In this context, a crucial element is the passage to the lower-dimensional Einstein frame. This is a non-trivial procedure because the Weyl rescaling needed to cast the action in the Einstein frame gives an additional contribution to the metric over metric variations. The final result is given by \cite{Palti:2024voy}
\begin{equation}\label{lowDimEinsteinAction}
 \begin{split}
  S_{d,grav}=\frac{1}{2}\int d^dx\,\,\sqrt{- g_{E}}\,&\bigl(\tilde R_{E}+e^{-2\tau+2D}\,\hat R_k-K_{\sigma\sigma}\,( \partial \,\sigma)^2\\
  &- K_{\tau\tau}\,( \partial\tau)^2-2K_{\Phi\,\tau}(\partial \Phi)(\partial\tau) -K_{\Phi\Phi}( \partial \Phi)^2\bigr),
 \end{split}
\end{equation}
where the Einstein frame metric is defined by the relation
\begin{equation}\label{EinsteinFrame}
 d s_{d}^2= e^{2D}\,ds_{E}^2= e^{2D+2\sigma}\,d\hat s_{E}^2\qquad \text{with} \qquad D=\frac{2\Phi-k\tau}{d-2}\,.
\end{equation}
 Following the notation of section \ref{AdSvolumevariations}, in the Einstein frame action we wrote $\tilde R_{E}=e^{-2\sigma}\hat R_{E}$.\footnote{The kinetic terms in \eqref{lowDimEinsteinAction} are expressed in terms of the Einstein frame metric, for example $( \partial \,\sigma)^2=g^{mn}_E\partial_m\sigma\partial_n\sigma$.} 
 
We will often refer to the combination $D$ as the lower-dimensional dilaton, and to $\Phi$ as the ten-dimensional dilaton. Note though that typically the lower dimensional dilaton is normalised without the $\frac{2}{d-2}$ prefactor in front of $\Phi$. Our conventions are convenient in that they ensure the kinetic term for $D$ always has the same (unit) normalisation in lower dimensions. 

Finally, the metric coefficients are given by \cite{Palti:2024voy}
\begin{equation}
\begin{split}\label{MetricCoefficients}
 &K_{\sigma\sigma}=-(d-1)(d-2)\,,\qquad K_{\tau\tau}=k^2\,\left(\frac{d-1}{d-2}\,-\frac{k-1}{k}\right)\,,\\ &K_{\Phi\Phi}=\frac{4}{d-2}\,,\qquad \qquad \qquad K_{\Phi\,\tau}=-\frac{2k}{d-2}\,.
 \end{split}
\end{equation}
We point out that \eqref{lowDimEinsteinAction} implies that the action of external and internal volume variations is just the sum of them. Moreover, we observe that the Weyl rescaling \eqref{EinsteinFrame} gave new contributions to the action, that are taken into account in \eqref{MetricCoefficients}.

Finally, we point out that choosing the more complicated variation \eqref{AdSvariationssigma1} with the auxialiary variation $\sigma_1$, the result would be again \eqref{lowDimEinsteinAction} plus the linear term $-2pe^{-2(\sigma+\sigma_1)+2D}z\,\partial_z\sigma_1$.

\subsubsection{Internal variations of product spaces}

A very common situation in string theory is when the internal space is a direct product of lower-dimensional manifolds. In what follows we extend the above results to vacua geometries of this type, so of the form
\begin{equation}
 \text{AdS}_d\times Y_{n_1}\times \cdots \times Y_{n_k}\,.
\end{equation}
For example, this is the case for the $\mrm{AdS}_3\times S^3\times \mathrm{CY}_2$ and $\mrm{AdS}_3\times S^3\times S^3\times S^1$ vacua, that we will consider in detail in section \ref{AdS3vacua}.

Let us consider the dimensional reduction of the higher-dimensional action \eqref{EHdilatonAction} over the internal manifold $Y_{n_1}\times \cdots \times Y_{n_k}$. As we did in simpler metric variations, we will suppose that volume factors of the spaces $Y_{n_i}$ are allowed to vary independently over AdS$_d$. For simplicity of notation, we outline the derivation for the case of $\text{AdS}_d\times Y_{n_1}\times Y_{n_2}$ vacua and then we will present the general formula.
We can consider the following metric deformation,
\begin{equation}
  ds_{d+n_1+n_2}^2=e^{2\sigma}d \hat s^2_d+e^{2\tau_1}\,d\hat s^2_{n_1}+e^{2\tau_2}\,d\hat s^2_{n_2}\,,
\end{equation}
where $\sigma$, $\tau_1$ and $\tau_2$ are functions over AdS$_d$. The factors $d\hat s^2_{n_1}$, $d\hat s^2_{n_2}$ are the metrics over unit radius $Y_{n_1}$, $Y_{n_2}$. The idea of the computation is just iterating the dimensional reductions over the spaces $Y_{n_i}$. For instance, starting with the partial reduction over $Y_2$, the action \eqref{EHdilatonAction} can be written as\footnote{This result can be obtained by applying the standard differential geometry formula $R= R_d+e^{-2\tau}\, R_k-k(k+1)\,  g_d^{mn} \partial_m \,\tau \partial_n\tau-2k\, \nabla_d^2\,\tau$ and integrating by part the Laplacian term. For more details, see \cite{Palti:2024voy}.}
\begin{equation}
   \begin{split}
   &S=
  \frac{1}{2\kappa^2}\int d^{d+n_1+n_2}x\sqrt{- g_{d+n_1}}\,\sqrt{\hat g_{n_2}}\,e^{k_2\,\tau_2-2\Phi}\,\bigl( R_{d+n_1}+e^{-2\tau_2}\,\hat R_{n_2}\\
  &+k_2(k_2-1)\, g_{d+n_1}^{mn} \partial_m \,\tau_2 \partial_n\tau_2 \nn-4k_2\, g_{d+n_1}^{mn}\partial_m \Phi\,\partial_n\tau_2+4g_{d+n_1}^{mn} \partial_m \Phi\, \partial_n\Phi\bigr)\,,
 \end{split}
\end{equation}
where, for simplicity of notation, we called $\kappa^{-2}$ the higher-dimensional gravitational coupling and we kept implicit external volume variations.
Adopting the same procedure with respect to $Y_{n_1}$ we obtain the result
\begin{equation}
   \begin{split}\label{dimRed2tau}
   S=&\frac{1}{2}\int d^{d}x\sqrt{- g_{d}}\,\,e^{k_1\tau_1+k_2\tau_2-2\Phi}\,\bigl( R_{d}+e^{-2\tau_1}\,\hat R_{n_1}+e^{-2\tau_2}\,\hat R_{n_2}+4g_{d}^{mn} \partial_m \Phi\, \partial_n\Phi\\
  &+k_1(k_1-1)\, g_{d}^{mn} \partial_m \,\tau_1 \partial_n\tau_1 \nn+k_2(k_2-1)\, g_{d}^{mn} \partial_m \,\tau_2 \partial_n\tau_2 \nn+2k_1k_2\,g_{d}^{mn} \partial_m \tau_1 \partial_n\tau_2 \nn\\
  &-4k_1\, g_{d}^{mn}\partial_m \Phi\,\partial_n\tau_1-4k_2\, g_{d}^{mn}\partial_m \Phi\,\partial_n\tau_2\bigr)\,,
 \end{split}
\end{equation}
where we fixed the higher-dimensional coupling as $\kappa^2=\text{Vol}(Y_{n_1})\text{Vol}(Y_{n_2})$ with $\text{Vol}(Y_{n_i})=\int d^{n_i}x\sqrt{\hat g_{n_i}}$.
We may now define the lower-dimensional Einstein frame as 
\begin{equation}\label{EinsteinFrame2tau}
 d s_{d}^2= e^{2D}\,ds_{E}^2= e^{2D-2\sigma}\,d\hat s_{E}^2\qquad \text{with} \qquad D=\frac{2\Phi-k_1\tau_1-k_2\tau_2}{d-2}\,.
\end{equation}
If we now include the contribution of external variations and we cast the action in the Einstein frame \eqref{dimRed2tau}, we obtain the following effective action
\begin{equation}
   \begin{split}\label{dimRed2tau}
  S_{d,grav}=&\frac{1}{2}\int d^dx\,\,\sqrt{- g_{E}}\,\bigl(\tilde R_{E}+e^{-2\tau_1+2D}\,\hat R_{n_1}+e^{-2\tau_2+2D}\,\hat R_{n_2}-K_{\sigma\sigma}\,( \partial \,\sigma)^2\\
  &- K_{\tau_1\tau_1}\,( \partial\tau_1)^2-K_{\tau_1\tau_1}\,( \partial\tau_1)^2- K_{\tau_1\tau_2}\,( \partial\tau_1)( \partial\tau_2)\\
  &-2K_{\Phi\,\tau_1}(\partial \Phi)(\partial\tau_1)-2K_{\Phi\,\tau_2}(\partial \Phi)(\partial\tau_2) -K_{\Phi\Phi}( \partial \Phi)^2\bigr)\,,
  \end{split}
\end{equation}
where the metric coefficients have the following form
\begin{equation}
\begin{split}\label{MetricCoefficientstau2}
 & K_{\tau_1\tau_1}=k_1^2\,\left(\frac{d-1}{d-2}\,-\frac{k_1-1}{k_1}\right)\,,\qquad K_{\tau_2\tau_2}=k_2^2\,\left(\frac{d-1}{d-2}\,-\frac{k_2-1}{k_2}\right)\,,\\ &K_{\Phi\Phi}=\frac{4}{d-2}\,,\qquad K_{\Phi\,\tau_1}=-\frac{2k_1}{d-2}\,,\qquad K_{\Phi\,\tau_2}=-\frac{2k_2}{d-2}\,,\qquad K_{\tau_1\,\tau_2}=\frac{k_1k_2}{d-2}\,,
 \end{split}
\end{equation}
where we recall that the negative contribution of external volume variation is given by $K_{\sigma\sigma}=-(d-1)(d-2)$.
Finally, one can iterate the procedure and extend this result to the general case of metric variations of $\text{AdS}_d\times Y_{n_1}\times \cdots \times Y_{n_k}$ spaces. The action is given by
\begin{equation}
   \begin{split}\label{dimRedGenera}
  S_{d,grav}=&\frac{1}{2}\int d^dx\,\,\sqrt{- g_{E}}\,\bigl(\tilde R_{E}-K_{\sigma\sigma}\,( \partial \,\sigma)^2-K_{\Phi\Phi}( \partial \Phi)^2\\
  &+\sum_{i,j}\biggl(e^{-2\tau_i+2D}\,\hat R_{n_i}- K_{\tau_i\tau_j}\,( \partial\tau_i)( \partial\tau_j)-2K_{\Phi\,\tau_i}(\partial \Phi)(\partial\tau_i) \biggr)\,,
  \end{split}
\end{equation}
where the Einstein frame is defined as $d s_{d}^2= e^{2D}\,ds_{E}^2$ with $D=\frac{2\Phi-\sum k_i\tau_i}{d-2}$. The metric coefficients are given by
\begin{equation}
\begin{split}\label{MetricCoefficientgeneral}
 & K_{\tau_i\tau_i}=k_i^2\,\left(\frac{d-1}{d-2}\,-\frac{k_i-1}{k_i}\right)\,,\quad K_{\tau_i\,\tau_j}=\frac{k_ik_j}{d-2}\,,\quad K_{\Phi\,\tau_i}=-\frac{2k_i}{d-2}\,,\quad K_{\Phi\Phi}=\frac{4}{d-2}\,.
 \end{split}
\end{equation}
We will employ this formula to derive metric variations of stringy AdS$_3\times S^3\times \mathrm{CY}_2$ and AdS$_3\times S^3\times S^3\times S^1$ backgrounds.

\subsection{Flux variations}
\label{Sec:fluxes}

Solutions in string theory, and M-theory, have fluxes turned on. Families of solutions are such that these flux parameters vary over the family. We therefore need to include their contribution to the metric and distance.

Let us consider Freund-Rubin vacua, as the flux configuration in these geometries is particularly simple. In fact, in these vacua the spacetime is a direct product AdS$_d\times Y_k$ and the flux entirely fills the directions of the external and/or internal space. Moreover, the ten-dimensional dilaton is trivial, $\Phi=0$. In this situation fluxes can be of two types:
\begin{itemize}
 \item Electric:$\qquad \hat F_d=d\hat C_p$\,,
 \item Magnetic: $\qquad \hat F_k=\star \hat F_d \qquad \text{with}\qquad  \hat F_d= d\hat C_p$\,,
\end{itemize}
where $\hat C_p$ is a gauge potential with $p=d-1$. Choosing Poincar\'e coordinates \eqref{AdSPoincarecord}, we can write the explicit expressions
\begin{equation}\label{gaugepot}
 \hat C_p=-z^{-p}\,dx^0\wedge \cdots \wedge dx^{p-1}\qquad \text{with}\qquad \hat F_d=(-1)^p\,p\,\text{vol}_d\,.
\end{equation}
In order to introduce flux variations, we need first to make explicit in the notation the physical scaling of the flux as 
\be 
F_d=e^{\alpha}\hat F_d \;,
\ee
where $\alpha$ is the flux parameter.
We can then introduce flux variations by promoting $\alpha$ to a function over AdS. This leads to the following relations
\begin{equation}\label{fluxvariationFR}
 C_p=e^\alpha\hat C_p\qquad \text{with} \qquad F_d=dC_p=(-1)^p\,e^{\alpha}\left(p-z\partial_z\alpha  \right)\text{vol}_d\,.
\end{equation}
The flux variation is off-shell, that is we cannot consider \eqref{fluxvariationFR} as a solution to the equations of motion. Moreover, the choice for the potential $C_p$ implies that only derivatives along $z$ contribute to the flux variation. In other words, with our prescription for flux variations, the AdS radial coordinate has a special role.

An important point is that electomagnetic duality between fluxes does not hold off-shell. In fact, the condition $\hat F_k=\star \hat F_d$ is an equation of motion in the democratic formulation. Once we go off-shell, we can vary the electric and magnetic parts of fluxes independently. This is important because the magnetic fluxes $F_k$ are evaluated over a compact space and so are quantized. On the other hand, electric $F_d$ fluxes can have non-trivial off-shell variations because they are defined over AdS, which is a non-compact manifold. Therefore, to evaluate the flux contributions we can always dualise appropriately to an electric frame, and then vary the electric flux. 

The above prescription then implies that we should evaluate the action for electric fluxes. So take
\begin{equation}\label{10dfluxaction}
S_{flux}=\frac{1}{2\kappa_{d+k}^2}\int d^{d+k}x\sqrt{-g}\,\bigl(- \frac12|F_d|^2\bigr)\,,
\end{equation}
with the flux variation \eqref{fluxvariationFR}. This leads to the following lower-dimensional off-shell action \cite{Li:2023gtt}
\begin{equation}
\begin{split}\label{fluxvariationAction}
 S_{d,flux}=&\frac12\,\int d^dx\sqrt{-g_E}\,e^{2\alpha-2p\sigma}\biggl(\frac12\,e^{2p\sigma-2pD}|\hat F_d|^2_E-\frac12(\partial \alpha)^2+pe^{2D-2(\sigma+\sigma_1)}z\partial_z\alpha\biggr)\,,
 \end{split}
\end{equation}
where with the subscript $E$ we indicate that the contractions are considered with respect to the Einstein frame metric $ds_E^2=e^{-2D}ds^2_d$. Specifically, the kinetic term of $\alpha$ can be defined naturally from the calculation as
\begin{equation}\label{kineticterm}
 (\partial\alpha)^2\equiv e^{-2(\sigma+\sigma_1)+2D}z^2(\partial_z\alpha)^2=g_E^{zz}(\partial_z\alpha)^2\,.
\end{equation}

Note that the overall sign of \eqref{fluxvariationAction} is opposite to the one resulting from the direct dimensional reduction of (\ref{10dfluxaction}). As discussed in detail in \cite{Li:2023gtt}, the dimensional reduction of electric fluxes leads to top-forms in $d$ dimensions. For this class of fluxes a boundary term is needed in order to reproduce a well-defined action \cite{Duff:1989ah,Groh:2012tf}. This boundary term switches the overall sign in front of the on-shell action. The off-shell action \eqref{fluxvariationAction} must be consistent with this feature otherwise it will not reproduce the right vacuum energy when evaluated on-shell.

The first term in the action (\ref{fluxvariationAction}) is the standard on-shell contribution of the flux to the vacuum energy. The second term is the kinetic term of flux variations \eqref{fluxvariationFR} and gives a contribution to the total metric over variations. The third term does not have a distance interpretation, but can be cancelled against the AdS variations with $\sigma_1$ introduced in \eqref{AdSvariationssigma1}, as explained in the next section.

As discussed above, for magnetic fluxes the variation must be taken over the dual electric flux. The off-shell action can be obtained from the standard action for magnetic fluxes $S_{flux}=\frac{1}{2\kappa_{d+k}^2}\int d^{d+k}x\sqrt{-g}\left(- \frac12|F_k|^2\right)$ and applying the following equalities\footnote{One has to use that $\star F_k=\star(\star F_d)=-(-1)^{kd}F_d$.}
\begin{equation}\label{idHodge}
 - F_k\,\wedge  \star F_k=(-1)^{kd}\,F_k\, \wedge\, F_d=F_d\,\wedge \,F_k=F_d\wedge \star F_d\,.
 \end{equation}
 From these relations it follows that the result for the off-shell action in $d$ dimensions have the same form as \eqref{fluxvariationAction}.

\subsection{The total metric over one-parameter vacua}

We have introduced all the required variations to define a metric over one-parameter vacua. We consider simple families with a single internal factor. The total action is the sum of the actions for metric and dilaton variations \eqref{offshellactionMetricSigma1} and \eqref{lowDimEinsteinAction}, and that of flux variations \eqref{fluxvariationAction}. This leads to the following result \cite{Li:2023gtt,Palti:2024voy}
\begin{equation}
\begin{split}\label{totalFRaction}
 S_{d}=&S_{d,grav}+S_{d,flux}=\\
 &\frac12\int d^dx\,\,\sqrt{- g_{d}}\bigl(\tilde R_{E}+e^{-2\tau+2D}\,\hat R_k+\frac12\,e^{2\alpha-2pD}|\hat F_d|^2_E\\
 &-K_{\sigma\sigma}\,( \partial \,\sigma)^2- K_{\tau\tau}\,( \partial\tau)^2-2K_{\Phi\,\tau}(\partial \Phi)(\partial\tau) -K_{\Phi\Phi}( \partial \Phi)^2-\frac12e^{2\alpha-2p\sigma}(\partial \alpha)^2\\
 &-2pe^{2D-2(\sigma+\sigma_1)}z\,\partial_z\sigma_1+pe^{2\alpha-2p\sigma+2D-2(\sigma+\sigma_1)}z\partial_z\alpha\bigr)\,,
 \end{split}
\end{equation}
where the metric coefficients are given in \eqref{MetricCoefficients}.
The supplementary AdS variation $\sigma_1$ is introduced in \eqref{AdSvariationssigma1}. Note that we restrict all these variations to the AdS radial direction only, so $\sigma=\sigma(z),\, \sigma_1=\sigma_1(z), \,\tau=\tau(z), \,\Phi=\Phi(z)$. This requirement is needed in order to keep the metric variations consistent with the flux variations.

As we explained in previous section, the equations of motion relate the different parameters on shell. We then need to implement these relations. The relations are solution-specific. A simple class of solutions are Freund-Rubin vacua, which have a single internal factor, and a constant dilaton. One then has
\begin{equation}\label{EOMFR}
 \sigma+\sigma_1=\tau\,,\qquad \qquad \alpha=p\,\sigma\,,\qquad \qquad \Phi=0\,,
\end{equation}
with $p=3,\,4,\,6$ for AdS$_4\times S^7$, AdS$_5\times S^5$, and AdS$_7\times S^4$. The flux is electric in the AdS$_4\times S^7$, magnetic in AdS$_7\times S^4$, and self-dual in AdS$_5\times S^5$.
Our strategy consists of imposing the above conditions when the variations are gauged. Crucially, the extra variation $\sigma_1$ is not fixed by the field equations. We can then choose it in a way so that the linear terms in \eqref{totalFRaction} cancel each other \cite{Li:2023gtt,Palti:2024voy}
\begin{equation}\label{sigma1CondFR}
 \sigma_1=\frac{\alpha}{2}\,.
\end{equation}
If we now solve together \eqref{EOMFR} and \eqref{sigma1CondFR}, we obtain a one-parameter family of solutions. We can then impose such conditions on the total action \eqref{totalFRaction} obtaining
\begin{equation}\label{actionmetricFRfinal}
  S_d=\frac12\int d^dx\,\sqrt{- g_{E}}\,\bigl(\tilde R_{E}+e^{-2\tau+2D}\,\hat R_k+\frac12\,e^{2\alpha-2pD}|\hat F_d|_E^2-K\,(\partial\sigma)^2 \bigr)\,,
\end{equation}
where $K$ is the total metric on the space of variations \cite{Li:2023gtt}
\begin{equation}\label{metricFR}
 K=-(d-1)(d-2)+\frac{k^2}{4}\,\left(\frac{d-1}{d-2}\,-\frac{k-1}{k}\right)\left(d+1\right)^2+\frac12(d-1)^2\,.
\end{equation}
For the specific cases AdS$_4\times S^7$, AdS$_5\times S^5$, and AdS$_7\times S^4$ one obtains the positive definite values\footnote{The case of $\text{AdS}_5\times S^5$ is subtle because of the self-duality of the flux. We refer to \cite{Li:2023gtt} for a discussion on that case.}
\begin{equation}\label{FRpositivemetric}
 K_{\text{AdS}_4\times S^7}=\frac{1563}{8}\,,\qquad K_{\text{AdS}_5\times S^5}=116\,,\qquad  K_{\text{AdS}_4\times S^7}=\frac{516}{5}.
\end{equation}
This analysis holds also in more complicated compactifications where the external and internal spaces are scale separated, and where there is a dilaton variation. This is the situation studied in \cite{Palti:2024voy}, where the case of so-called DGKT vacua \cite{DeWolfe:2005uu} was considered. These are $\ma N=1$ AdS$_4$ vacua obtained by Calabi-Yau compactification with orientifolds. The crucial property of this family of vacua is to be described by a single parameter, defined by the magnetic 4-flux within the Calabi-Yau. In other words, even if these compactification are much more complicated that the Freund-Rubin ones, they admit a notion of metric which is very simple, depending only on one single variation. We refer to \cite{Palti:2024voy} for a detailed derivation of the metric over the variations DGKT vacua, which basically follow the same procedure as the one outlined above. The final result obtained in \cite{Palti:2024voy} for the total metric over field variations in DGKT vacua is given by the (positive) number
\begin{equation}\label{DGKTpositivemetric}
 K_{\text{DGKT}}=\frac{3376}{27}\,.
\end{equation}
This completes our analysis of one-parameter families of vacua, and we now consider multi-parameter families.

\section{The metric over type IIA $\mathrm{AdS}_4\times \mathbb{CP}^3$ vacua}
\label{sec:ads4cp3}

In this section we compute the metric over Type IIA $\mathrm{AdS}_4\times \mathbb{CP}^3$ vacua. These ten-dimensional geometries describe the weakly coupled limit of M-theory vacua $\mathrm{AdS}_4\times S^7/\mathbb{Z}_k$. From the perspective of the metric computation, dealing with these vacua is more involved than the Freund-Rubin case. Specifically, they have two sets of independent fluxes, $F_2$ and $F_4$, implying that the space of parameters of the corresponding supergravity solution is two-dimensional.

We begin by reviewing the main properties of this family of type IIA vacua, discussing the on-shell conditions and the derivation of the vacuum energy in the four-dimensional theory. We then introduce flux variations and compute the corresponding off-shell action. Finally, we obtain the metric over vacua variations, which is a $2\times 2$ matrix. We find that the coefficients of this matrix are constant, so that it is flat, and that it has positive eigenvalues. In the last part of the section we discuss how the conditions over fluxes that make these AdS solutions trustable can be used to constrain the two-dimensional space of variations.

\subsection{The on-shell solutions}
\label{sec:onshellAdS4CP3}

Type IIA $\mathrm{AdS}_4\times \mathbb{CP}^3$ vacua \cite{Watamura:1983hj, Nilsson:1984bj} arise from M-theory vacua $\mathrm{AdS}_4\times S^7/\mathbb{Z}_k$ that are associated to stacks of M2 branes on a $\mathbb{C}^4/\mathbb{Z}_k$ singularity \cite{Aharony:2008ug}. From a geometric perspective, the explicit ten-dimensionl background can be obtained by writing the $S^7/\mathbb{Z}_k$ manifold as a circle fibration over $\mathbb{CP}^3$
and dimensionally reducing over the circle. The metric and fluxes can be written in the {\itshape string frame} as \cite{Watamura:1983hj, Nilsson:1984bj}
\begin{equation}
 \begin{split}\label{AdS4CP3}
 &ds^2_{10}=e^{2\sigma}\left(d\hat s_4^2 +4 d\hat s_6^2 \right)\,,\\
 &F_2=4e^{\sigma-\Phi}\,J\,,\\
 &F_4=-3e^{3\sigma-\Phi}\,\text{vol}_4\,,\\
 \end{split}
\end{equation}
where $d\hat s_4^2$ and $d\hat s_6^2$ are respectively the metrics of unit radius AdS$_4$ and $\mathbb{CP}^3$. The electric 4-flux is a Freund-Rubin flux along the AdS$_4$ directions and $F_2$ is a magnetic flux, which is turned on along the Kahler form $J$ of $\mathbb{CP}^3$. We refer to appendix \ref{app:F2fluxes} for more details on the $\mathbb{CP}^3$ geometry.
The above vacua constitute a two-parameter family of solutions of type IIA supergravity. In our notation the two constant parameters of the solution are associated to the scalings in the two parameters $\sigma$ and $\Phi$, with the latter being the ten-dimensional dilaton.

Since we look at independent off-shell variations of the geometry and fluxes, we can cast the the expressions in \eqref{AdS4CP3} in a fashion which is more suitable for this purpose.
The vacuum geometry is described by a family of {\itshape string frame} metrics of the following type
\begin{equation}
\begin{split}\label{AdS4CP4metricgeneral}
 &ds^2_{10}=e^{2\sigma}d\hat s_4^2 +e^{2\tau} d\hat s_6^2 \qquad \text{with}\qquad d\hat s^2_4=\frac{1}{z^2}\left(ds^2_{M_3}+e^{2\sigma_1}dz^2\right)\,,
 \end{split}
\end{equation}
where the parameters $\sigma$ and $\tau$ describe external and internal volume variations as in section \ref{Sec:internalvolume}. In the above expression we wrote the AdS$_4$ metric as in \eqref{AdSvariationssigma1}, with the auxialiary parameter $\sigma_1$.
We recall that when $\sigma$ and $\sigma_1$ are constants, the above four-dimensional metric can be cast in the standard AdS$_4$ metric form with shifted radius $e^{\sigma+\sigma_1}$ (through a change of variable $z\rightarrow e^{\sigma_1}z$). As we observed previously, the inclusion of $\sigma_1$ is crucial to reproduce a well-defined notion of metric over off-shell variations.

What makes these solutions more interesting, compared to Freund-Rubin vacua, is their dependence on two different fluxes. Specifically, we can introduce two constant parameters associated to the physical scaling of $F_2$ and $F_4$ fluxes. As we argued in section \ref{Sec:fluxes}, flux variations are non-trivial when they are defined over electric fluxes, since they extend along the AdS directions, that are non-compact and are allowed to vary off-shell. We thus focus on the electric fluxes in AdS$_4\times \mathbb{CP}^3$ vacua and write them in terms of their physical scaling parameters, denoted as $\alpha$ and $\beta$,
\begin{equation}
\begin{split}\label{AdS4CP4fluxgeneral}
 &F_4=-3\,e^\alpha\,\text{vol}_4\,,\\
 & F_8=-3\,e^\beta\text{vol}_4\wedge J\wedge J\qquad \text{with}\qquad F_2=\star_{10}\,F_8=6\,e^{\beta-4\sigma-\sigma_1-2\tau}\,J\,,
 \end{split}
\end{equation}
where the Hodge-dualization leading to $F_2$ is explained in appendix \ref{app:F2fluxes}.

Given the above family of metrics and fluxes, the type IIA equations of motion can be written as a set of algebraic conditions among the parameters. 
Specifically, one has the following on-shell relations for $\mathrm{AdS}_4\times \mathbb{CP}^3$ vacua
\begin{equation}\label{onshell}
 \sigma+\sigma_1=\tau-\log2\,,\qquad \alpha=3\sigma-\Phi\,,\qquad \beta=3\sigma +4\tau-\Phi-\log6\,,
\end{equation}
where from the last two relations we can write $\beta=\alpha+4\tau-\log6$. The vacuum energy can then be extracted by evaluating the ten-dimensional action over the internal space. The four-dimensional action can be obtained by evaluating the IIA action over the metric \eqref{AdS4CP4metricgeneral} and fluxes \eqref{AdS4CP4fluxgeneral}, and then imposing the on-shell conditions \eqref{onshell}. The relevant contributions from the type IIA action are the following
\begin{equation}
 S_{\text{II}A}=\frac{1}{2\kappa_{10}^2}\,\int d^{10}x\sqrt{-g}\,\left(e^{-2\Phi}R -\frac12\,|F_4|^2-\frac12|F_2|^2\right)+\cdots\,.
\end{equation}
We use the same notations and conventions of sections \ref{Sec:internalvolume} and \ref{Sec:fluxes}. In particular, the four-dimensional Einstein frame metric is defined as
\be 
 ds^2_E=e^{-2D}ds^2_4 \;\;,
 \label{cp3einsmetf}
 \ee 
 with $d\hat s^2_4$ the metric of unit radius AdS$_4$ and 
 \be 
 D=\Phi-3\tau \;.
 \ee  
 After some manipulations we obtain the folllowing four-dimensional action
\begin{equation}
\begin{split}\label{AdS4CP3action}
 S_4=\frac12\int d^4x\sqrt{-g_E}\,&\biggl(\tilde R_E+48 \,e^{-2\tau+2D}-\frac92\,e^{2\alpha-8\sigma-2\sigma_1+2\Phi+2D}\\
 & -54\,e^{2\beta-8\sigma-2\sigma_1-8\tau+2\Phi+2D}\biggr)\,.
 \end{split}
\end{equation}
The sum of the last three terms gives rise to the four-dimensional vacuum energy, once they are evaluated on the on-shell conditions \eqref{onshell}. Specifically, the second term is associated to the Ricci scalar of unit radius $\mathbb{CP}^3$, that is $\hat R_6=48$ (see \eqref{lowDimEinsteinAction} for the general derivation of this contribution). The third and the last contributions respectively belong to 2- and 4-fluxes. The derivation of the $F_4$ contribution is the same as that of the Freund-Rubin vacua AdS$_4\times S^7$, with the trivial inclusion of the dilaton scaling in the passage to the Einstein frame. We refer to section 3.1 in \cite{Li:2023gtt} for more details.\footnote{\label{foot:boundary}Importantly, in order to reproduce the right value of the four-dimensional vacuum energy, the derivation of the on-shell action in four dimensions \eqref{AdS4CP3action} requires a boundary term associated to the electric $F_4$ flux \cite{Duff:1989ah,Groh:2012tf}. The effect of this term is switching the overall sign of the $|F_4|^2$ term in the on-shell action. For more details see the discussion in section 3.1 in \cite{Li:2023gtt}.} Finally, the contribution to the action of the $F_2$ flux is derived in appendix \ref{app:F2fluxes} (see, in particular, equations \eqref{app:F2action1} and \eqref{app:F2action2}).

Now, imposing the on-shell conditions \eqref{onshell} on the action \eqref{AdS4CP3action}, we obtain the four-dimensional vacuum energy,
\begin{equation}
 -2\Lambda\equiv 6e^{-2(\sigma+\sigma_1)+2D}\,,
\end{equation}
where the right-hand side is the result of the sum of the last three contributions.
From this relation it is particularly manifest that the AdS$_4$ square radius in the four-dimensional theory is given by $e^{2(\sigma+\sigma_1)+2D}$. The presence of the four-dimensional dilaton $D$ is standard and it fixes the normalization of the AdS radius in the four-dimensional Einstein frame. 


\subsection{The variations of $F_4$ and $F_8$ fluxes}

In this section we calculate the flux variations for $\mathrm{AdS}_4\times \mathbb{CP}^3$ vacua, using the results of section \ref{Sec:fluxes}. Let us start with the electric flux $F_4$. As we explained after equations \eqref{gaugepot}, the flux variation for a purely electric Freund-Rubin flux in four dimension has the following form
\begin{equation}
 C_3=-e^\alpha z^{-3}\,dx^0\wedge dx^1 \wedge dx^{2}\qquad \text{with} \qquad F_4=dC_3=-\,e^{\alpha}\left(3-z\partial_z\alpha  \right)\text{vol}_4\,,
\end{equation}
where $\alpha$ is an infinitesimal deformation of the flux parameter defined locally along the AdS radial coordinate.
The derivation of the off-shell action for these variations in performed in section 3.2 in \cite{Li:2023gtt} in the case of AdS$_4\times S^7$. The only difference here is the additional scaling in the ten-dimensional dilaton appearing in the passage to the Einstein frame. In particular, we obtain the following result
\begin{equation}\label{F4action10D}
 S_{F_4}=-\frac{1}{4\kappa_{10}^2}\int d^{10}x\sqrt{-g}|F_4|^2=\frac{1}{2}\int d^{4}x\sqrt{-g_4}\,e^{2\alpha-8\sigma-2\sigma_1+6\tau}(3-z\partial_z\alpha)^2\,,
\end{equation}
where we have fixed the ten-dimensional gravitational coupling as $\kappa_{10}^2=\text{Vol}(\mathbb{CP}^3)=\int  d^6y\,\sqrt{\hat g_6}$. We refer to appendix \ref{app:F2fluxes} for some useful formulas.

We can go now to the four-dimensional Einstein frame (\ref{cp3einsmetf}). We thus find the following four-dimensional action
\begin{equation}
\begin{split}\label{F4action}
 S_{4,F_4}=\frac{1}{2}\int d^{4}x\sqrt{-g_E}\,e^{2\alpha-6\sigma+2\Phi}\left(-\frac92e^{-2(\sigma+\sigma_1)+2D}-\frac12(\partial\alpha)^2+3e^{-2(\sigma+\sigma_1)+2D}z\partial_z\alpha\right).
 \end{split}
\end{equation}
The kinetic term of $\alpha$ has been written in its covariant form, absorbing the various scalings to obtain the inverse metric $g_E^{zz}$ as we did in \eqref{kineticterm}, so with $(\partial\alpha)^2\equiv e^{-2(\sigma+\sigma_1)+2D}z^2(\partial_z\alpha)^2=g_E^{zz}(\partial_z\alpha)^2$.
An important comment is about the overall sign of \eqref{F4action}. As we mentioned in footnote \ref{foot:boundary}, the dimensional reduction of electric Freund-Rubin fluxes leads to top-forms in lower dimensions. For this class of fluxes a boundary term is needed in order to reproduce a well-defined action in four dimensions \cite{Duff:1989ah,Groh:2012tf}. The effect of this boundary term is to switch the overall sign in front of the lower-dimensional on-shell action. The off-shell action \eqref{F4action} must be consistent with this feature otherwise we cannot reproduce the right vacuum energy when we evaluate the action on-shell. For this reason, the expressions \eqref{F4action10D} and \eqref{F4action} differ by an overall sign.\footnote{For more details we refer to section 3.2 in \cite{Li:2023gtt}.}

Let us now consider the two-form flux $F_2$. This is a magnetic flux, so the off-shell variation must be taken over the electric dual $F_8$. Specifically, we can write
\begin{equation}
\begin{split}\label{C7andF8}
& C_7=-e^\beta z^{-3}\,dx^0\wedge dx^1 \wedge dx^{2}\wedge J\wedge J\,,\\
&F_8=dC_7=-\,e^{\beta}\left(3-z\partial_z\beta  \right)\text{vol}_4\wedge J\wedge J\,,
\end{split}
\end{equation}
where $\beta$ is now a function along the radial coordinate of AdS$_4$. As shown in \eqref{idHodge}, the contribution to the action of the $F_8$ flux can be obtained from the $F_2$ action by applying fundamental relations of Hodge duality. We refer to appendix \ref{app:F2fluxes} for the explicit derivation of the $F_8$ action (see \eqref{app:F8action1} and \eqref{app:F8action2}). Taking the above flux variations one obtains the following off-shell action
\begin{equation}\label{F2action10D}
 S_{F_8}=\frac{1}{2\kappa_{10}^2}\int d^{10}x\sqrt{-g}|F_8|^2=-\frac{1}{2}\int d^{4}x\sqrt{-g_4}\,e^{2\beta-8\sigma-2\sigma_1-2\tau}6(3-z\partial_z\beta)^2\,,
\end{equation}
where again we have fixed the ten-dimensional gravitational coupling as $\kappa_{10}^2=\text{Vol}(\mathbb{CP}^3)$.
Going to the four-dimensional Einstein frame (\ref{cp3einsmetf}), we obtain the following action
\begin{equation}
\begin{split}\label{F2action}
 S_{4,F_8}=\frac{1}{2}\int d^{4}x\sqrt{-g_E}\,&e^{2\beta-6\sigma-8\tau+2\Phi}\biggl(-54e^{-2(\sigma+\sigma_1)+2D}\\
 &-6(\partial\beta)^2+36e^{-2(\sigma+\sigma_1)+2D}z\partial_z\beta\biggr)\,,
 \end{split}
\end{equation}
where $(\partial\beta)^2\equiv e^{-2(\sigma+\sigma_1)+2D}z^2(\partial_z\beta)^2=g_E^{zz}(\partial_z\beta)^2$.
We are now ready to compute the total metric over vacua variations.

\subsection{The total metric over $\mrm{AdS}_4\times \mathbb{CP}^3$ vacua}

In the previous section we introduced the off-shell variations associated to the scalings in $\sigma, \sigma_1, \tau, \Phi, \alpha, \beta$. We calculated the contributions to the action of each variation separately. In this section we combine them. 
The total type IIA action is
\begin{equation}
\begin{split}\label{IIAterms}
 &S_{\text{IIA}}=S_{\text{EH}}+S_{\Phi}+S_{F_4}+S_{F_2}+\cdots\,,\\
  &S_{grav}=S_{\text{EH}}+S_{\Phi}=\frac{1}{2\kappa_{10}^2}\,\int d^{10}x\sqrt{-g}\,e^{-2\Phi}\left(R+4g^{mn}\partial_m\Phi\partial_n\Phi   \right)\,,\\
  &S_{flux}=-\frac{1}{2\kappa_{10}^2}\,\int d^{10}x\sqrt{-g}\,\left(\frac12\,|F_4|^2-\frac12|F_8|^2\right) \,.\\
  \end{split}
  \end{equation}
  The metric and dilaton variations, coming from the action $S_{grav}$, can be obtained from the general formula $\eqref{lowDimEinsteinAction}$, with the particular choice of $d=4$ external and $k=6$ internal dimensions. The flux part of the action corresponds to summing up the actions \eqref{F4action} and \eqref{F2action}, computed in previous section. 

Combining these results, we thus obtain the total four-dimensional off-shell action
\begin{equation}
\begin{split}\label{totalActionAdS4CP3}
 S_{4}&=\,S_{4,grav}+S_{4,flux}=\\
 &=\frac12\int d^4x\,\,\sqrt{- g_{E}}\biggl[\tilde R_{E}+48 \,e^{-2\tau+2D}+6\,( \partial \,\sigma)^2-24\,( \partial\tau)^2+12(\partial \Phi)(\partial\tau)-2\,( \partial \Phi)^2\\
 &-6e^{-2(\sigma+\sigma_1)+2D}z\partial_z\sigma_1+e^{2\alpha-6\sigma+2\Phi}\biggl(-\frac92e^{-2(\sigma+\sigma_1)+2D}-\frac12(\partial\alpha)^2+3e^{-2(\sigma+\sigma_1)+2D}z\partial_z\alpha\biggr)\\
 &+e^{2\beta-6\sigma-8\tau+2\Phi}\left(-54e^{-2(\sigma+\sigma_1)+2D}-6(\partial\beta)^2+36e^{-2(\sigma+\sigma_1)+2D}z\partial_z\beta\right)\biggr]\,.
 \end{split}
\end{equation}

In analogy to the case of Freund-Rubin vacua, we can now require that the metric variation associated to $\sigma_1$ is related to flux variations $\alpha$ and $\beta$. In particular, the condition
\begin{equation}
 \sigma_1=\frac{1}{6}\left(3\alpha+\beta\right) \;,
 \label{sigm1cp3sol}
\end{equation}
implies that the linear terms in derivatives cancel. 

The remaining on-shell relations (\ref{onshell}) can be written in terms of $\Phi$ and $\sigma$ as
\begin{equation}
\label{finalvariationsAdS4CP3}
 \tau=9\sigma-2\Phi+\frac12\log\left(\frac{32}{3} \right) \;\;, \;\;\;
 \alpha=3\sigma-\Phi\;\;,\;\;\;
 \beta=39\sigma-9\Phi+3\log\left(\frac{8}{3} \right)\;.
 \end{equation}
 We observe that the above family depends on two independent off-shell variations, consistently with the number of independent parameters of AdS$_4\times \mathbb{CP}^3$ vacua.

 Imposing the conditions \eqref{finalvariationsAdS4CP3} on the off-shell action \eqref{totalActionAdS4CP3}, we obtain the following result
\begin{equation}
\label{finalActionAdS4CP3}
 S_{4}=\frac12\int d^4x\,\,\sqrt{- g_{E}}\biggl[\tilde R_{E}-2\Lambda-K_{\mathbb{CP}^3}\,( \partial \,\Sigma)^2\biggr]\;.
\end{equation}
Here we wrote the constant contributions in terms of the four-dimensional vacuum energy $\Lambda=-3e^{-2(\sigma+\sigma_1)+2D}$. The kinetic terms for the variations are written in terms of a two-component vector $\Sigma$, and a $2\times 2$ matrix $K_{\mathbb{CP}^3}$, with 
\begin{equation}\label{metricAdS4CP3}
K_{\mathbb{CP}^3}=
\begin{pmatrix}
2196 & -546 \\
-546 & 136
\end{pmatrix}
\qquad \text{,} \qquad
\Sigma= \begin{pmatrix}
\sigma \\
\Phi
\end{pmatrix}\,.
\end{equation}
The above metric is positive definite, it has eigenvalues $\lambda_i$ approximately given by
\be 
\lambda_1 \sim 2332 \;,\;\; \lambda_2 \sim 0.2 \;.
\ee
It therefore yields a well defined notion of distance over the space of solutions. 

\subsection{Physical paths and geodesics in solution space}

In (\ref{metricAdS4CP3}), we have chosen to parametrize the solution using the scalings in $\sigma$ and $\Phi$ (the AdS radius and the dilaton). We could have chosen some other parameterisation of the space of solutions. This would not affect the positivity of the metric. However, each parameterisation has a physical meaning in that the range of parameters is not arbitrary. For example, we cannot reliably extend the dilaton to strong coupling $\Phi \rightarrow 0$. Similarly, we cannot trust high curvature regimes, say $\sigma \rightarrow 0$.

In terms of $\sigma$ and $\Phi$ we can therefore consider paths, or families of solutions, with 
\begin{equation}\label{preholographic}
\sigma \gg 1 \qquad\text{and} \qquad  \Phi \ll -1\;\;.
\end{equation}
The on-shell solutions are fixed in terms of the flux choices. The type IIA solution \eqref{AdS4CP3} involves $n$ units of $F_4$ on AdS$_4$ and $k$ units of $F_2$ flux over the 2 cycle within the $\mathbb{CP}^3$. Their relation to $\sigma$ and $\Phi$ is \cite{Aharony:2008ug}
\begin{equation}\label{changeAdS4CP3}
 e^{2\sigma}\sim\sqrt{\frac{n}{k}}\qquad\text{and} \qquad e^{2\Phi}\sim\sqrt{\frac{n}{k^5}}\;\;.
\end{equation}
From the above expression we can see that there are flux choices which can cover the full $\sigma$ and $\Phi$ space subject to the conditions (\ref{preholographic}), that is
\begin{equation}\label{holographic}
n \gg k \qquad\text{and} \qquad  k^5 \gg n\;\;.
\end{equation}

Indeed, since the metric is flat we can parameterise geodesics by straight lines. The physically reliable space of paths, or families of solutions, is most conveniently parameterised by first changing to $\log n$ and $\log k$ space, with associated metric $K_{\mathbb{CP}^3,\mathrm{flux}}$. Using \eqref{changeAdS4CP3}, we can change variables to that space 
\begin{equation}
\begin{split}
 S_{4}=\frac12\int d^4x\,\,\sqrt{- g_{E}}\biggl[\tilde R_{E}-2\Lambda-\ma K_{\mathbb{CP}^3,\mathrm{flux}}\,( \partial \ma F)^2\biggr]\,,
 \end{split}
\end{equation}
where the total metric $\ma K_{\mathbb{CP}^3,\mathrm{flux}}$ is given by
\begin{equation}
\ma K_{\mathbb{CP}^3,\mathrm{flux}}=
\begin{pmatrix}
\frac{17}{2} & 25 \\
25 & \frac{155}{2}
\end{pmatrix}
\qquad \text{with} \qquad
\ma F= \begin{pmatrix}
 \log n \\
 \log k
\end{pmatrix}\,.
\end{equation}

Geodesics are straight lines in $\log n$ and $\log k$ space, and so can be parametrised by $\gamma$ with  
\be 
n = k^{\gamma} \;.
\ee
Then infinite distance paths are such that $k \rightarrow \infty$, and a choice of $\gamma$ within the range
\be 
\label{gammarange}
1 \leq \gamma \leq 5 \;.
\ee 
The lower bound in (\ref{gammarange}) corresponds to the requirement to stay at weak coupling, while the upper bound is required to stay at weak curvature. The distance, as a function of $k$, along any infinite distance geodesic, so a given choice of $\gamma$, can be evaluate by changing to $k$ and $\gamma$ variables as
\begin{equation}
\begin{split}
 &S_{4}=\frac12\int d^4x\,\,\sqrt{- g_{4}}\biggl[\tilde R_{E}-2\Lambda-\ma K_{\gamma}\,( \partial \log k)^2\biggr]\,,\\
 &\ma K_{\mathbb{CP}^3,\gamma}=\frac12(17+100\gamma+155\gamma^2)\,.
 \end{split}
 \end{equation}
The metric $K_{\mathbb{CP}^3,\gamma}$ then gives the distance along any physical infinite distance geodesic path in the space of solutions.

\section{Weyl rescalings versus Weyl variations and moduli}
\label{sec:weylmoduli}

In cases where there are multiple parameter families of solutions, there appears the possibility of moduli. Moduli are parameters in the family of solution which leave the fluxes invariant. In other words, they have no flux (or curvature) induced potential, they are flat directions. One way to define a modulus is as a parameter which does not change the energy of the vacuum. In terms of AdS solutions, it means that the cosmological constant in the lower-dimensional Einstein frame does not change when a modulus is varied. 

The existence of moduli within the family of solutions introduces a certain ambiguity to the distance calculation prescription. In this section we describe this ambiguity, and explain how to resolve it. This will yield the general procedure for dealing with moduli, and we will utilise this approach in the specific solutions of sections \ref{AdS3vacua} and \ref{sec:ads3s3s3s1}.

Let us recall that there are two types of contributions to the distance coming from variations of the metric. They are both related to the Weyl rescaling formula. Recall that if we have two metrics in $d$ dimensions,  
\be 
ds^2_d = e^{2\sigma} d\hat{s}^2_d \;,
\label{weylsigmafr}
\ee 
then the Ricci scalars of the two metrics are related as 
\be 
R_d = e^{-2\sigma} \left[ \hat{R}_d - \left(d-1\right)\left(d-2\right)(\hat{\partial}\sigma )^2 - 2 \left(d-1\right)\hat{\nabla}^2_d \;\sigma \right] \;,
\label{weylformula}
\ee 
where the hats on the derivatives denote that the indices are contracted with the metric $d\hat{s}^2_d$. 

The derivative terms in (\ref{weylformula}) form a contribution to the distance. They therefore arise either from changing the metric in the action, a {\it Weyl rescaling}, for example when going to the Einstein frame. Or they can arise while keeping the metric the same, but promoting $\sigma$ from a constant parameter in the metric to one which has some spatial dependence, which we refer to as a {\it Weyl variation}. The contribution from $\sigma$ in (\ref{offshellActionExternalVariations}) is a Weyl variation contribution, while for example $K_{\tau\tau}$ in (\ref{MetricCoefficients}) includes contributions from Weyl rescalings. 

When we calculate the contribution from a Weyl variation, rather than a Weyl rescaling, the metric stays the same. We are simply promoting the factor in front of the metric to have some spatial dependence. There is therefore a certain ambiguity which arises: for what metric do we calculate the Weyl variation? To understand this, consider taking (\ref{weylsigmafr}), and relating it further to a metric which undergoes two Weyl rescalings, by parameters $D_1$ and $D_2$, so as
\be 
ds^2_E =  e^{2\sigma} \;e^{-2D_1} \; e^{-2D_2}\; d\hat{s}^2_d   \;. 
\ee 
We have denoted the final metric $ds^2_E$ as an Einstein frame metric, since this is how it will arise in the examples, but for now we can consider it as any metric. 

Now if we write our action in terms of the Einstein frame metric $ds^2_E\;$, the Weyl rescalings by $D_1$ and $D_2$ will contribute some quadratic variation terms as in (\ref{weylformula}), so will contribute to the distance. We can also have a contribution from a Weyl variation taken with respect to $\sigma$. This is how things are calculated in sections \ref{sec:metricoverads} and \ref{sec:ads4cp3}, with $D_1+D_2=D$, the lower-dimensional dilaton. 

The important point is that because the Weyl variation does not modify the metric, we could have considered taking the variation with respect to say $\sigma-D_1$ instead of $\sigma$. In other words, we could have performed a Weyl rescaling by $D_1$, and then performed the variation. Or we could have first performed the variation, and only then Weyl rescaled. The two results are not equivalent. More precisely, varying only $\sigma$ gives a contribution as in (\ref{offshellActionExternalVariations}), so of
\be 
 (d-1)(d-2)(\partial \,\sigma)^2 \;,
\ee 
while varying $\sigma-D_1$ would instead give
\be 
 (d-1)(d-2)\big(\partial \left(\sigma-D_1\right)\big)^2 \;.
\ee 

This ambiguity is resolved by noting that since both a Weyl rescaling and a Weyl variation contribute to the distance, we should not vary with respect to $\sigma - D_1$ since the contribution from $D_1$ is already accounted for by performing the Weyl rescaling. We would be double counting the contribution this way. This is why we vary only with respect to $\sigma$ in sections \ref{sec:metricoverads} and \ref{sec:ads4cp3}. 

However, in the presence of moduli there is a new subtlety. Moduli are parameters which keep the cosmological constant fixed (in the Einstein frame). This means that their contribution to the AdS distance should be zero. Double-counting zero is allowed of course, and so when we take a Weyl variation, which contributes only to the AdS distance, we should be able to do it with respect to $\sigma$ or with respect to $\sigma - r\chi$, where $\chi$ is a modulus parameter and $r$ is some constant. 

What fixes the appropriate variation choice? Well we know that moduli should not give any contributions to the distance from Weyl variations, only from Weyl rescalings. Therefore, we should adopt the following prescription. Consider solving for $\sigma$, the external metric scale factor, in terms of the other solution parameters using the on-shell relations. After doing so, we can write $\sigma$ as 
\be 
\sigma = \sigma_f + r \;\chi \;,
\ee 
where $\chi$ is a modulus parameter, $r$ is an appropriate constant, and $\sigma_{f}$ depends only on the flux parameters and not on the modulus (or moduli). Now instead of performing a Weyl variation with respect to $\sigma$, we perform the variation with respect to $\sigma_f$. A way to think of this is to consider performing a Weyl rescaling by $D_1 = r \chi$, performing the variation, so with respect to $\sigma - D_1$. And then performing a second Weyl rescaling to complete the required rescaling to go to the lower-dimensional Einstein frame, so with $D_2 = D - r \chi$. The result is equivalent to performing a Weyl variation of only $\sigma_f$.

The result of this prescription is that the contributions of the modulus to the distance come only from the Weyl rescalings, and not from the Weyl variation. This indeed makes physical sense: a modulus variation by definition does not vary the AdS radius, and so should not contribute to the AdS distance. 

We will apply this prescription explicitly in sections \ref{AdS3vacua} and \ref{sec:ads3s3s3s1}. But it holds completely generally, for any number of moduli. It means that moduli and AdS variations factorise and should be treated as independent.  

\section{The metric over type IIB $\mathrm{AdS}_3\times S^3\times \mathrm{CY}_2$ vacua}
\label{AdS3vacua} 

In this section we compute the metric over families of type IIB AdS$_3\times S^3\times \mathrm{CY}_2$ vacua, with the $\mathrm{CY}_2$ being $K3$ or $T^4$.\footnote{For a non-exhaustive list of references on these vacua we refer to \cite{Gibbons:1993sv,Gibbons:1994vm,Strominger:1997eq,Maldacena:1997re,Giveon:1998ns}.} These type of vacua add a layer of complexity over the $\mathrm{AdS}_4\times \mathbb{CP}^3$ case studied in section \ref{sec:ads4cp3}. They also have a two-parameter family of fluxes, but they further have a (one-dimensional) moduli space. A modulus corresponds to a continuous parameter of the solution space which is not fixed by the equations of motion in the presence of the fluxes. In other words, the geometry and the fluxes do not induce a potential for the modulus direction, and it is therefore often referred to as a {\itshape flat direction}.

As discussed in section \ref{sec:weylmoduli}, the presence of a modulus requires a certain modification of the usual prescription for evaluating the distance along the space of solutions. The point is that Weyl variations should not contribute to the modulus distance, only to the AdS distance. In practice, it means we only perform a Weyl variation of the flux part of $\sigma$, denoted $\sigma_f$ in section \ref{sec:weylmoduli}. The modulus distance instead receives contributions only from Weyl rescaling, and these will reproduce known results for it, see for example \cite{Gukov:2004ym}. 

\subsection{The on-shell solutions}
\label{sec:onshellAdS3S3T4}

In our notation the on-shell metric and fluxes can be written as follows
\begin{equation}
 \begin{split}\label{AdS3S3T4}
 &ds^2_{10}=e^{2\sigma}\left(d\hat s_3^2 + d\hat s_{S^3}^2\right) +e^{2\tau_2}d\hat s^2_{\text{CY}_2}\,,\\
&F_3=2e^{2\sigma-\Phi}\,\text{vol}_3+2e^{2\sigma-\Phi}\,\text{vol}_{S^3}\,,
 \end{split}
\end{equation}
where $d\hat s_3^2$, $d\hat s_{S^3}^2$, $d\hat s^2_{\text{CY}_2}$ are the metrics of unit radius AdS$_3$, $S^3$ and $\text{CY}_2$. The RR 3-flux has one electric component along the AdS$_3$ and a magnetic one along the $S^3$.
The above vacua belong to the following family of geometries
\begin{equation}
 \begin{split}\label{generalAdS3S3T4}
 &ds^2_{10}=e^{2\sigma}d\hat s_3^2 + e^{2\tau_1}d\hat s_{S^3}^2 +e^{2\tau_2}d\hat s^2_{\text{CY}_2}\qquad \text{with}\qquad d\hat s^2_3=\frac{1}{z^2}\left(ds^2_{M_2}+e^{2\sigma_1}dz^2\right)\,,\\
 &F_3=F_3^{\text{e}}+F_3^{\text{m}}\,,\\
 &F_3^{\text{e}}=2e^{\alpha}\,\text{vol}_3\,,\qquad \qquad  F_3^{\text{m}}=-\star_{10} F_7^{\text{e}}=2e^{\beta-3\sigma-\sigma_1+3\tau_1-4\tau_2}\,\text{vol}_{S^3}\,,
 \end{split}
\end{equation}
where we split the electric and magnetic terms of the $F_3$ flux and introduced again the auxialiary variation $\sigma_1$ in the parametrization of AdS$_3$. The electric fluxes  $F_3^{\text{e}}$ and $F_7^{\text{e}}$ can be defined by gauge potentials of the following form
\begin{equation}
\begin{split}
&C_2=-e^\alpha z^{-2}\,dx^0\wedge dx^1\qquad \text{with} \qquad F_3^{\text{e}}=2e^{\alpha}\,\text{vol}_3\,,\\
& C_6=-e^\beta z^{-2}\,dx^0\wedge dx^1\wedge \text{vol}_{\mathrm{CY}_2}\qquad \text{with} \qquad F_7^{\text{e}}=dC_6=2e^\beta\text{vol}_3\wedge \text{vol}_{\mathrm{CY}_2}\,,
\end{split}
\end{equation}
where $\text{vol}_{\mathrm{CY}_2}$ is the volume form of the unit radius manifold CY$_2$.
We can now compute the equations of motion of type IIB supergravity, obtaining the following set of on-shell conditions
\begin{equation}\label{onshellAdS3S3T4}
 \tau_1=\sigma+\sigma_1\,,\qquad \alpha=2\sigma-\Phi\,,\qquad \beta=2\sigma+4\tau_2-\Phi\,.
\end{equation}

This a system of three equations for seven parameters. Further, the parameter $\sigma_1$ is fixed, as usual, to cancel the linear terms in the variations as in (\ref{mixedtermsAdS3S3T4}). We therefore have three free parameters which specify the family of solutions. Two of these parameters are flux choices, while the third is a modulus. It is important to separate these two types of parameters, and we do so in section \ref{sec:ads3s3cy2modver}. 

Let us first calculate the three-dimensional vacuum energy. This follows from evaluating the type IIB action over the family of metrics and fluxes \eqref{generalAdS3S3T4} and imposing the on-shell conditions (\ref{onshellAdS3S3T4}).
The relevant contributions from the type IIB action are the following
\begin{equation}
 S_{\text{II}B}=\frac{1}{2\kappa_{10}^2}\,\int d^{10}x\sqrt{-g}\,\left(e^{-2\Phi}R -\frac12\,|F_3^\text{e}|^2-\frac12|F_3^\text{m}|^2\right)+\cdots\,.
\end{equation}
Evaluating this action over the vacua geometries \eqref{generalAdS3S3T4} we obtain the following three-dimensional action
\begin{equation}
\begin{split}\label{AdS3S3T4action}
 S_3=\frac12\int d^3x\sqrt{-g_E}\,&\biggl(\tilde R_E+6 \,e^{-2\tau_1+2D}-2\,e^{2\alpha-6\sigma-2\sigma_1+2\Phi+2D}\\
 & -2\,e^{2\beta-6\sigma-2\sigma_1-8\tau_2+2\Phi+2D}\biggr)\,,
 \end{split}
\end{equation}
where we fixed the type IIB gravitational coupling as $\kappa^2_{10}=\text{Vol}(S^3)\text{Vol}(\text{CY}_2)$. The Einstein frame metric is defined as 
\be 
ds^2_E=e^{-2D}ds^2_3\;,
\ee
where the three-dimensional dilaton obeys the general relation \eqref{EinsteinFrame2tau}, which in this particular case has the form 
\be 
D=2\Phi-3\tau_1-4\tau_2 \;.
\label{3ddilads3cy2}
\ee 
Imposing the on-shell conditions \eqref{onshellAdS3S3T4}, the sum of the last three terms leads to the three-dimensional vacuum energy. In particular, the second term is associated to the Ricci scalar of the $S^3$, namely $\hat R_3=6$, and the last two terms respectively belong to electric and magnetic components of the $F_3$ flux.\footnote{As for Freund-Rubin AdS$_4\times S^7$ and AdS$_4\times \mathbb{CP}^3$ vacua, the reduction of the electric flux $F^{\text{e}}_3$ generates a top-form in three dimensions. For this reason, we included a boundary term that switched the sign in front of the corresponding on-shell action. This contribution must be included in order to reproduce the right value of the three-dimensional vacuum energy. For more details, see  \cite{Duff:1989ah,Groh:2012tf,Li:2023gtt}.\label{ft7}}
 We thus obtain
\begin{equation}
 \Lambda = -e^{-2(\sigma+\sigma_1)+2D}\,,
 \label{3dlambdaads3s3y2}
\end{equation}
that is the precise value of the vacuum energy in the three-dimensional Einstein frame.

\subsection{Flux variations}

Let us introduce flux variations in AdS$_3\times S^3\times \text{CY}_2$ vacua. From the expressions of on-shell fluxes in \eqref{generalAdS3S3T4} we observe that we can introduce relevant variations over the electric fluxes $F_3^{\text{e}}$ and $F_7^{\text{e}}$. As usual, we may start from the gauge potentials $C_2$ and $C_6$ as follows
\begin{equation}
\begin{split}\label{FluxvariationsAdS3S3T4}
 &C_2=-e^\alpha z^{-2}\,dx^0\wedge dx^1 \qquad \text{with} \qquad F_3^\text{e}=dC_2=\,e^{\alpha}\left(2-z\partial_z\alpha  \right)\text{vol}_3\,,\\
 &C_6=-e^\beta z^{-2}\,dx^0\wedge dx^1\wedge\text{vol}_{\text{CY}_2} \quad \text{with} \quad F_7^\text{e}=dC_6=e^{\beta}\left(2-z\partial_z\beta  \right)\text{vol}_3\wedge \text{vol}_{\text{CY}_2}\,,
\end{split}
\end{equation}
where now $\alpha$ and $\beta$ are functions of $z$. We recall that the above configurations do not satisfy anymore the type IIB equations of motion. We can explicitly compute the off-shell action for the above variations
\begin{equation}\label{F3actionsAdS3S3T4}
\begin{split}
 &S_{F^\text{e}_3}=-\frac{1}{4\kappa_{10}^2}\int d^{10}x\sqrt{-g}|F^\text{e}_3|^2=\frac{1}{4}\int d^{3}x\sqrt{-g_3}\,e^{2\alpha-6\sigma-2\sigma_1+3\tau_1+4\tau_2}(2-z\partial_z\alpha)^2\,,\\
 &S_{F^\text{e}_7}=\frac{1}{4\kappa_{10}^2}\int d^{10}x\sqrt{-g}|F^\text{e}_7|^2=-\frac{1}{4}\int d^{3}x\sqrt{-g_3}\,e^{2\beta-6\sigma-2\sigma_1+3\tau_1-4\tau_2}(2-z\partial_z\beta)^2\,.
 \end{split}
\end{equation}
In the Einstein frame, we obtain the following contributions to the lower-dimensional effective action
\begin{equation}\label{F3actionsAdS3S3T4EF}
\begin{split}
 &S_{3, F^\text{e}_3}=\frac{1}{2}\int d^{3}x\sqrt{-g_E}\,e^{2\alpha-4\sigma+2\Phi}\bigl(-2\,e^{-2(\sigma+  \sigma_1)+2D}-\frac12(\partial\alpha)^2+2e^{-2(\sigma+\sigma_1)+2D}z\partial_z\alpha\bigr),\\
 &S_{3,F^\text{e}_7}=\frac{1}{2}\int d^{3}x\sqrt{-g_E}\,e^{2\beta-4\sigma-8\tau_2+2\Phi}\bigl(-2\,e^{-2(\sigma+\sigma_1)+2D}-\frac12(\partial\beta)^2+2e^{-2(\sigma+\sigma_1)+2D}z\partial_z\beta\bigr)\,.
 \end{split}
\end{equation}
Note that we absorbed the scaling $e^{-2(\sigma+\sigma_1)+2D}$ into the kinetic terms of $\alpha$ and $\beta$, as in \eqref{kineticterm}. We notice that when $\alpha$ and $\beta$ are evaluated to constant values we recover the corresponding on-shell contributions to the vacuum energy \eqref{AdS3S3T4action}. (This includes the flip of the overall sign for the $F^\text{e}_3$ flux action in (\ref{F3actionsAdS3S3T4EF}), as discussed in footnote \ref{ft7}.)

\subsection{The full variations over $\mrm{AdS}_3\times S^3\times \mathrm{CY}_2$ vacua}\label{sec:metricAdSS3T4}

Let us now combine the results to write the full quadratic variation of the action with respect to all the parameters in (\ref{generalAdS3S3T4}), that is $ \sigma, \sigma_1, \tau_1, \tau_2, \Phi, \alpha, \beta$.
The relevant terms of the type IIB action are
\begin{equation}
\begin{split}\label{IIBterms-offshell}
 &S_{\text{IIB}}=S_{\text{EH}}+S_{\Phi}+S_{F_3}+S_{F_7}+\cdots\,,\\
  &S_{grav}=S_{\text{EH}}+S_{\Phi}=\frac{1}{2\kappa_{10}^2}\,\int d^{10}x\sqrt{-g}\,e^{-2\Phi}\left(R+4g^{mn}\partial_m\Phi\partial_n\Phi   \right)\,,\\
  &S_{flux}=-\frac{1}{2\kappa_{10}^2}\,\int d^{10}x\sqrt{-g}\,\left(\frac12\,|F_3|^2-\frac12|F_7|^2\right) \,.\\
  \end{split}
  \end{equation}
  The action for metric variations has been derived in a general form in \eqref{dimRed2tau}, with the metric coefficients given in \eqref{MetricCoefficientstau2}. The action $S_{grav}$ is a particular case of the general formula \eqref{dimRed2tau}. Specifically, setting the parameters $d=3$, $k_1=3$, $k_2=4$, we obtain the following three-dimensional off-shell action for metric and dilaton variations
  \begin{equation}
   \begin{split}\label{offshellActionMetricAdS3S3T4}
  S_{3,grav}=\frac{1}{2}\int d^3x\,\,\sqrt{- g_{E}}&\,\bigl(\tilde R_{E}+6\,e^{-2\tau_1+2D}-6e^{-2(\sigma+\sigma_1)+2D}z\partial_z\sigma_1\\
  &+2( \partial \,\sigma_f)^2-3\,( \partial\tau_1)^2-4\,( \partial\tau_2)^2 -( \partial D)^2\bigr)\,,
  \end{split}
\end{equation}
where we expressed the variations in terms of the three-dimensional dilaton (\ref{3ddilads3cy2}). We also performed the Weyl variation with respect to $\sigma_f$ only, rather than $\sigma$. That is, we removed the modulus contribution to this variation, as explained in section \ref{sec:weylmoduli}. The explicit relation between $\sigma$ and $\sigma_f$ is given in (\ref{pcfinalvariationsAdS3S3T4}). 

Combining the above action with the actions of flux variations \eqref{F3actionsAdS3S3T4EF}, we get the following expression:
\begin{equation}
\begin{split}\label{totalActionAdS3S3T4}
 &S_{3}=\,S_{3,grav}+S_{3,flux}=\frac12\int d^3x\,\,\sqrt{- g_{E}}\,\biggl[\tilde R_{E}+6 \,e^{-2\tau_1+2D}+2( \partial \,\sigma_f)^2-3\,( \partial\tau_1)^2\\
 &-4\,( \partial\tau_2)^2-( \partial D)^2-6e^{-2(\sigma+\sigma_1)+2D}z\partial_z\sigma_1+e^{2\alpha-4\sigma+3\tau_1+D+4\tau_2}\bigl(-2\,e^{-2(\sigma+\sigma_1)+2D}\\
 &-\frac12(\partial\alpha)^2+2e^{-2(\sigma+\sigma_1)+2D}z\partial_z\alpha\bigr)+e^{2\beta-4\sigma+3\tau_1+D-4\tau_2}\bigl(-2\,e^{-2(\sigma+\sigma_1)+2D}\\
 &-\frac12(\partial\beta)^2+2e^{-2(\sigma+\sigma_1)+2D}z\partial_z\beta\bigr)\biggr]\,.
 \end{split}
\end{equation}
As usual, the parameter $\sigma_1$ needs to chosen so that the action variations are purely quadratic, so to cancel the linear variations. This fixes
\begin{equation}\label{mixedtermsAdS3S3T4}
 \sigma_1=\frac{1}{3}(\alpha+\beta)\,.
\end{equation}
Using this, we obtain the quadratic action
\begin{equation}
\begin{split}\label{quadtotalActionAdS3S3T4}
 S_{3}=\frac12\int d^3x\,\,\sqrt{- g_{E}}\,\biggl[ \;&\tilde R_{E}+6 \,e^{-2\tau_1+2D}+2( \partial \,\sigma_f)^2-3\,( \partial\tau_1)^2 
 -4\,( \partial\tau_2)^2-( \partial D)^2
 \\ & +e^{2\alpha-4\sigma+3\tau_1+D+4\tau_2}\bigl(-2\,e^{-2(\sigma+\sigma_1)+2D}-\frac12(\partial\alpha)^2\bigr)\\
 &+e^{2\beta-4\sigma+3\tau_1+D-4\tau_2}\bigl(-2\,e^{-2(\sigma+\sigma_1)+2D}-\frac12(\partial\beta)^2\bigr)\;\biggr]\,.
 \end{split}
\end{equation}
At this point, we would normally directly extract the metric over the variations by imposing the relations (\ref{onshellAdS3S3T4}). However, due to the presence of the modulus in this background, we need to perform the procedure discussed in section \ref{sec:weylmoduli}. That is, we need to identify the relation between $\sigma$ and $\sigma_f$. We return to this in section \ref{sec:meadfghj}, but first we consider purely modulus variations.

\subsection{The metric over modulus variations}
\label{sec:ads3s3cy2modver}

Let us first consider the modulus variations. From the on-shell relations (\ref{onshellAdS3S3T4}), we can identify the modulus as a common shift by a parameter $\lambda$ which keeps the flux parameters $\alpha$ and $\beta$ fixed
\be 
\sigma \rightarrow \sigma + \lambda \;,\;\; \tau_1 \rightarrow \tau_1 + \lambda \;,\;\; \Phi \rightarrow \Phi + 2 \lambda \;.
\label{modlamshift}
\ee 
The modulus is a combination of $\tau_1$ and $\Phi$, or more conveniently, of $\tau_1$ and $D$. To fix this combination we first choose a combination $\psi$ which is invariant under the shift (\ref{modlamshift}). The other direction, the modulus $\chi$, is then taken as orthogonal to $\psi$, with respect to the metric induced by the quadratic action (\ref{quadtotalActionAdS3S3T4}). So we have
\bea
\psi &=& \tau_1 - D \;,\nn \\   
\chi &=& \frac34\tau_1 + \frac14 D \;.
\eea
The normalisation of $\chi$ is chosen so that it transforms as the modulus, so as $\lambda$ under (\ref{modlamshift}). For convenience, we also give the inverse relations
\bea 
\tau_1 &=& \chi+\frac14 \psi  \;,\nn \\   
D &=& \chi -\frac34 \psi \;.
\label{tau1Dtopc}
\eea 
Using these, the quadratic variations in the action (\ref{quadtotalActionAdS3S3T4}) are then written as
\be 
-3(\partial \tau_1)^2 - (\partial D)^2  = -4 (\partial \chi )^2 - \frac34 (\partial \psi )^2 \;\;.
\label{quadvarconvpt}
\ee
The kinetic terms for the modulus $\chi$ the matches the appropriate Einstein frame value. So the metric over modulus variations is as expected, and of course positive
\be 
K_{\mathrm{CY}_2,\chi\chi} = 4 \;.
\ee 

The modulus then factorises from the AdS variations. That is, the kinetic terms for $\chi$ do not mix with those of $\tau_2$ and $\psi$. Further, the Einstein frame vacuum energy (\ref{3dlambdaads3s3y2}) is independent of $\chi$, as expected since it is a flat direction
\be 
 \Lambda = -e^{-2(\sigma+\sigma_1)+2D} = -e^{-2\left( \tau_1 -D \right)} = -e^{-2\psi} \;,
 \label{Lambdindch}
\ee   
where we used the on-shell relations (\ref{onshellAdS3S3T4}).

We have therefore reproduced the expected behaviour under the moduli space variations. We now go on to consider the AdS variations.

\subsection{The metric over AdS variations}
\label{sec:meadfghj}

To extract the metric over AdS variations, we should impose the on-shell relations (\ref{onshellAdS3S3T4}) on the variation in the quadratic action (\ref{quadtotalActionAdS3S3T4}). However, as discussed in section \ref{sec:weylmoduli}, the presence of the modulus means we need to split the modulus and flux contributions to $\sigma$ first. Let us first write the on-shell relations in terms of $\psi$ and $\chi$ instead of $\tau_1$ and $D$, so using (\ref{tau1Dtopc}). We have
\begin{equation}
\label{pcfinalvariationsAdS3S3T4}
\sigma=\sigma_f  + \chi \,,\qquad \sigma_f = \frac14 (\alpha+\beta)\,,\qquad 
\psi=\frac73\left(\alpha+\beta\right) \,,
\qquad \tau_2=-\frac14\left(\alpha-\beta  \right) \;.
 \end{equation}
We should then impose these on the quadratic variations, (\ref{quadtotalActionAdS3S3T4}) and (\ref{quadvarconvpt}),  which read\footnote{Note that we have dropped the modulus kinetic term $\left( \partial \chi \right)^2$, since we are interested only in the AdS variations.}  
\bea
\label{pcquadtotalActionAdS3S3T4}
2( \partial \,\sigma_f)^2 
 -4\,( \partial\tau_2)^2 - \frac34 (\partial \psi )^2  
 -\frac12e^{2\alpha-4\sigma+4\chi+4\tau_2}(\partial\alpha)^2
 -\frac12e^{2\beta-4\sigma+4\chi-4\tau_2}(\partial\beta)^2\,. 
\eea
This yields the quadratic AdS variations, parameterised in terms of $\alpha$ and $\beta$, of
\be 
-\frac{113}{24} (\partial \alpha )^2 - \frac{89}{12} (\partial \alpha )(\partial \beta) -\frac{113}{24} (\partial \beta )^2 \;.
\ee 
The metric over the two-parameter space of AdS variations is therefore
\be 
K_{\mathrm{CY}_2}=
\left(
\begin{array}{cc}
	\frac{113}{24} & \frac{89}{24}  \\
	\frac{89}{24} & \frac{113}{24} 
	\end{array}
\right) \;.
\ee 
As in the other cases, the metric is constant and positive definite, with eigenvalues
\be 
\lambda_1 = \frac{101}{12}\;,\;\; \lambda_2 = 1\;.
\ee 

\section{The metric over type IIB $\mrm{AdS}_3\times S^3\times S^3\times S^1$ vacua}
\label{sec:ads3s3s3s1}

In this section we consider the case of $\mrm{AdS}_3\times S^3\times S^3\times S^1$ vacua.\footnote{For a non-exhaustive list of references on these vacua we refer to \cite{Cowdall:1998bu,Boonstra:1998yu,Gauntlett:1998kc,Gukov:2004ym}.} As mentioned above, these solutions are much more involved than the $\mrm{AdS}_3\times S^3\times \text{CY}_2$ ones. Specifically, in these vacuum configurations, there are two magnetic terms in the 3-flux that contribute non-trivially to the vacuum energy. This implies that variations of the external volume are non-linearly related to the volume of the three-spheres. To address this issue, we will express the variations in a specific form where all non-linear terms are associated with variations that do not exhibit an infinite-distance limit. This approach will allow us to simplify the problem and compute the metric following the same method used in previous cases.

\subsection{The on-shell solutions}

Let us start by summarizing the $\mrm{AdS}_3\times S^3\times S^3\times S^1$ solutions in type IIB supergravity. These backgrounds are described by a family of metrics and fluxes of the following type
\begin{equation}
 \begin{split}\label{AdS3S3S3S1}
 &ds^2_{10}=e^{2\sigma}d\hat s_3^2 +e^{2\tau_1} d\hat s_{S_1^3}^2 +e^{2\tau_2}d\hat s^2_{S_2^3}+e^{2\tau_3}d\hat s^2_{S^1}\,,\\
&F_3=2e^{2\sigma-\Phi}\,\text{vol}_3+2e^{2\tau_1-\Phi}\,\text{vol}_{S^3_1}-2e^{2\tau_2-\Phi}\,\text{vol}_{S^3_2}\,,
 \end{split}
\end{equation}
where $d\hat s_3^2$, $d\hat s_{S^3_1}^2$, $d\hat s^2_{S_2^3}$, $d\hat s^2_{S^1}$ are the metrics of AdS$_3$, $S^3_1$, $S^3_2$, $S^1$ with unit radius. The RR three-flux has one electric component along the AdS$_3$ directions and two magnetic ones along the three-spheres.
To be a solution to the equations of motion, the backgrounds \eqref{AdS3S3S3S1} have to satisfy the following non-linear relation between metric variations
\begin{equation}\label{AdS3S3S3S1solution}
 \sigma=\tau_1+\tau_2-\frac12\log\left(e^{2\tau_1}+e^{2\tau_2}\right)\,.
\end{equation}

As in the previous cases, we can express the above vacuum geometries keeping all possible variations independent,
\begin{equation}
 \begin{split}\label{generalAdS3S3S3S1}
 &ds^2_{10}=e^{2\sigma}d\hat s_3^2 +e^{2\tau_1} d\hat s_{S_1^3}^2 +e^{2\tau_2}ds^2_{S_2^3}+e^{2\tau_3}ds^2_{S^1}\quad \text{with}\quad d\hat s^2_3=\frac{1}{z^2}\left(ds^2_{M_2}+e^{2\sigma_1}dz^2\right)\,,\\
 &F_3=F_3^{\text{e}}+F_3^{\text{m}}\,,\\
 &F_3^{\text{e}}=2e^{\alpha}\,\text{vol}_3\,,\\
 &F_3^{\text{m}}=-\star_{10} F_7^{\text{e}}=2e^{\beta_1-3\sigma-\sigma_1+3\tau_1-3\tau_2-\tau_3}\,\text{vol}_{S^3_1}-2e^{\beta_2-3\sigma-\sigma_1-3\tau_1+3\tau_2-\tau_3}\,\text{vol}_{S^3_2}\,,
 \end{split}
\end{equation}
where we introduced the auxialiary variation $\sigma_1$ within the AdS$_3$ geometry. The electric flux $F_7^{\text{e}}$ can be defined by the following gauge potentials
\begin{equation}
\begin{split}
 &C_6=-e^{\beta_2} z^{-2}\,dx^0\wedge dx^1\wedge \text{vol}_{S^3_1}\wedge d\psi-e^{\beta_1} z^{-2}\,dx^0\wedge dx^1\wedge \text{vol}_{S^3_2}\wedge d\psi\,\\
 &F_7^{\text{e}}=dC_6=2e^{\beta_2}\text{vol}_3\wedge \text{vol}_{S^3_1}\wedge d\psi+2e^{\beta_1}\text{vol}_3\wedge \text{vol}_{S^3_2}\wedge d\psi\,,
 \end{split}
\end{equation}
where the coordinate $\psi$ parametrizes the circle $S^1$. In general, we have the following fundamental quantities: $\sigma, \tau_1, \tau_2, \tau_3, \Phi, \alpha, \beta_1, \beta_2$. Substituing the prescription \eqref{generalAdS3S3S3S1} into the type IIB equations of motion we obtain the on-shell conditions:
\begin{equation}
\begin{split}\label{onshellAdS3S3S3S1}
 &\sigma+\sigma_1=\tau_1+\tau_2-\frac12\log\left(e^{2\tau_1}+e^{2\tau_2}\right)\,,\qquad \qquad \alpha=2\sigma-\Phi\,,\\
 &\beta_1=3\sigma+\sigma_1-\tau_1+3\tau_2+\tau_3-\Phi\,,\qquad \qquad \beta_2=3\sigma+\sigma_1+3\tau_1-\tau_2+\tau_3-\Phi\,.
 \end{split}
\end{equation}
We stress that the variations in these equations are related non-linearly to each other. 

Let us compute the vacuum energy. The relevant terms in the ten-dimensional action are the following
\begin{equation}
 S_{\text{IIB}}=\frac{1}{2\kappa_{10}^2}\,\int d^{10}x\sqrt{-g}\,\left(e^{-2\Phi}R -\frac12\,|F_3^\text{e}|^2-\frac12|F_3^\text{m}|^2\right)+\cdots\,.
\end{equation}
Evaluating the type IIB action over the vacua $\mrm{AdS}_3\times S^3\times S^3\times S^1$ in \eqref{generalAdS3S3S3S1}, we obtain the following three-dimensional action
\begin{equation}
\begin{split}\label{AdS3S3T4action}
 S_3=\frac12\int & d^3x\sqrt{-g_E}\,\biggl(\tilde R_E+6 \,e^{-2\tau_1+2D}+6 \,e^{-2\tau_2+2D}-2\,e^{2(\alpha-3\sigma-\sigma_1+\Phi)+2D}\\
 & -2\,e^{2(\beta_1-3\sigma-\sigma_1-3\tau_2-\tau_3+\Phi)+2D}-2\,e^{2(\beta_2-3\sigma-\sigma_1-3\tau_1-\tau_3+\Phi)+2D}\biggr)\,,
 \end{split}
\end{equation}
where the Einstein frame metric is defined as 
\be 
ds^2_E=e^{-2D}ds^2_3 \;.
\ee 
The three-dimensional dilaton obeys the general relation \eqref{EinsteinFrame2tau}, which in this case has the form 
\be 
D=2\Phi-3\tau_1-3\tau_2-\tau_3 \;.
\ee 
The various contributions to the vacuum energy in (\ref{AdS3S3T4action}) are as follows. The second and the third action terms belong to the Ricci scalars of the 3-spheres. The third term is associated to the reduction of the electric component $F^{\text{e}}_3$.\footnote{In the reduction of the electric flux $F^{\text{e}}_3$ we included a boundary term that switched the sign in front of the corresponding on-shell action. See the previous and analogous reductions of top-forms in this work. For more details, see  \cite{Duff:1989ah,Groh:2012tf,Li:2023gtt}.} These contributions can be computed in a similar form to the other cases considered previously.

Let us now we impose the on-shell conditions \eqref{onshellAdS3S3S3S1} on the above action. The sum of the five contributions in the action precisely reproduces the right scaling and coefficient of the three-dimensional vacuum energy, which is defined as
\begin{equation}
 \Lambda\equiv-e^{-2(\sigma+\sigma_1)+2D}\,.
\end{equation}

Consider the last two terms in (\ref{AdS3S3T4action}). These correspond to the magnetic components of the three-form flux and behave in a peculiar way. We can isolate these contributions:
\begin{equation}\label{F3magVacuum}
 -\frac{1}{4\kappa_{10}^2}\,\int d^{10}x\sqrt{-g}\,|F_3^\text{m}|^2 \, \rightarrow \,\frac12\int d^3x\sqrt{-g_E}\,e^{2D}\left(-2e^{-2\tau_1}-2e^{-2\tau_2} \right) \,,
\end{equation}
where we imposed the on-shell conditions for $\beta_1, \beta_2$ in  \eqref{onshellAdS3S3S3S1} and we put the arrow to highlight the passage to the Einstein frame. In the other vacua considered, each flux term gains the typical on-shell scaling of the vacuum energy independently from each others, in our notation $e^{-2(\sigma+\sigma_1)+2D}$. Moreover, to get such a scaling behavior, it is usually enough to impose the equations of motion for the corresponding flux (the equations for $\beta_1, \beta_2$ in this case). In this sense, each action term identifies a contribution to the vacuum energy.
In this present case instead, the two magnetic contributions to the three-form flux combine together to give a single contribution. In fact, the expression \eqref{F3magVacuum} needs an additional on-shell condition to reproduce the right scaling behavior of the vacuum energy, specifically the equation linking $\tau_1, \tau_2$ and $\sigma$. Imposing such a relation on \eqref{F3magVacuum}, we obtain just a single contribution,
\begin{equation}\label{F3magVacuum2}
\begin{split}
 &\frac12\int d^3x\sqrt{-g_E}\,e^{2D}\left(-2e^{-2\tau_1}-2e^{-2\tau_2} \right)= \,\frac12\int d^3x\sqrt{-g_E}\,e^{-2(\sigma+\sigma_1)+2D}(-2) \,,\\
 &\text{with} \qquad \sigma+\sigma_1=\tau_1+\tau_2-\frac12\log\left(e^{2\tau_1}+e^{2\tau_2}\right)\,.
\end{split}
\end{equation}
As we are going to see, this peculiarity of the magnetic flux in AdS$_3\times S^3 \times S^3 \times S^1$ has important consequences for the prescription for flux variations.

\subsection{Flux variations}

Let us consider off-shell flux variations in AdS$_3\times S^3\times S^3\times S^1$ vacua. The flux configuration \eqref{AdS3S3S3S1} has electric and magnetic components for the three-form flux. We begin by introducing electric flux variations. These behave as in the AdS$_3\times S^3\times \mrm{CY}_2$ case, that is
\begin{equation}
\begin{split}
 &C_2=-e^\alpha z^{-2}\,dx^0\wedge dx^1 \qquad \text{with} \qquad F_3^\text{e}=dC_2=\,e^{\alpha}\left(2-z\partial_z\alpha  \right)\text{vol}_3\,.\\
\end{split}
\end{equation}
The derivation of the action for variations of $F_3^{e}$ is analogous to \eqref{FluxvariationsAdS3S3T4}, with the exception of the scalings in the internal volumes. We obtain the following expression
\begin{equation}\label{F3actionsAdS3S3S3S1}
\begin{split}
 &S_{F^\text{e}_3}=-\frac{1}{4\kappa_{10}^2}\int d^{10}x\sqrt{-g}|F^\text{e}_3|^2=\frac{1}{4}\int d^{3}x\sqrt{-g_3}\,e^{2\alpha-6\sigma-2\sigma_1+3\tau_1+3\tau_2+\tau_3}(2-z\partial_z\alpha)^2,
 \end{split}
\end{equation}
where we fixed the ten-dimensional coupling as $\kappa_{10}^2=\text{Vol}(S_1^3)\,\text{Vol}(S_2^3)\,\text{Vol}(S^1)$. We can express the above integral in the Einstein frame, obtaining
\begin{equation}\label{F3actionAdS3S3S3S1EF}
\begin{split}
 &S_{3, F^\text{e}_3}=\frac{1}{2}\int d^{3}x\sqrt{-g_E}\,\bigl(-2\,e^{-2(\sigma+\sigma_1)+2D}-\frac12(\partial\alpha)^2+2e^{-2(\sigma+\sigma_1)+2D}z\partial_z\alpha\bigr),
 \end{split}
\end{equation}
where we used the fundamental relation $\alpha=2\sigma-\Phi$ from \eqref{onshellAdS3S3S3S1} to cancel the overall scaling factor in \eqref{F3actionsAdS3S3S3S1}. As in previous cases, the overall sign of (\ref{F3actionAdS3S3S3S1EF}) is flipped due to the boundary term, relative to direct dimensional reduction.

We can now study flux variations for magnetic fluxes. To this aim, we can introduce $C_6$ gauge potentials of the following form
\begin{equation}
\begin{split}\label{C6variations}
 &C_6=-e^{\beta_2} z^{-2}\,dx^0\wedge dx^1\wedge \text{vol}_{S^3_1}\wedge d\psi-e^{\beta_1} z^{-2}\,dx^0\wedge dx^1\wedge \text{vol}_{S^3_2}\wedge d\psi\,,\\
 & F_7^\text{e}=dC_6=e^{\beta_2}\left(2-z\partial_z\beta_2\right)\text{vol}_3\wedge \text{vol}_{S^3_1}\wedge d\psi+e^{\beta_1}\left(2-z\partial_z\beta_1  \right)\text{vol}_3\wedge \text{vol}_{S^3_2}\wedge d\psi\,.
\end{split}
\end{equation}
The off-shell action for variations of $F_7^{e}$ has the following expression
\begin{equation}\label{F7actionsAdS3S3S3S1}
\begin{split}
 S_{F^\text{e}_7}=\frac{1}{4\kappa_{10}^2}\int d^{10}x\sqrt{-g}|F^\text{e}_7|^2=&-\frac{1}{4}\int d^{3}x\sqrt{-g_3}\,e^{2\beta_1-6\sigma-2\sigma_1+3\tau_1-3\tau_2-\tau_3}(2-z\partial_z\beta_1)^2\\
 &-\frac{1}{4}\int d^{3}x\sqrt{-g_3}\,e^{2\beta_2-6\sigma-2\sigma_1-3\tau_1+3\tau_2-\tau_3}(2-z\partial_z\beta_2)^2 \;.
 \end{split}
\end{equation}
Expressing the above action in the Einstein frame, we obtain the following result
\begin{equation}
\begin{split}\label{F7actionAdS3S3S3S1EF}
 &S_{3,F^\text{e}_7}=\\
 &\frac{1}{2}\int d^{3}x\sqrt{-g_E}\,\biggl[e^{2\beta_1-4\sigma-6\tau_2-2\tau_3+2\Phi}\bigl(-2\,e^{-2(\sigma+\sigma_1)+2D}-\frac12(\partial\beta_1)^2+2e^{-2(\sigma+\sigma_1)+2D}z\partial_z\beta_1\bigr)\\
 &+e^{2\beta_2-4\sigma-6\tau_1-2\tau_3+2\Phi}\bigl(-2\,e^{-2(\sigma+\sigma_1)+2D}-\frac12(\partial\beta_2)^2+2e^{-2(\sigma+\sigma_1)+2D}z\partial_z\beta_2\bigr)\biggr].
 \end{split}
\end{equation}

It is important to point out that the kinetic terms for $\beta_2, \beta_2$, as well as the linear terms, have two different scaling behaviors. This can be seen imposing the on-shell conditions for $\beta_1, \beta_2$, written in \eqref{onshellAdS3S3S3S1}. 
The flux action \eqref{F7actionAdS3S3S3S1EF} takes then the following form
\begin{equation}
\begin{split}\label{F7actionAdS3S3S3S1EF2}
 S_{3,F^\text{e}_7}=\frac{1}{2}\int d^{3}x\sqrt{-g_E}\,&\biggl[-2\,e^{-2(\sigma+\sigma_1)+2D}-\frac12e^{2(\sigma+\sigma_1)}\left(e^{-2\tau_1}(\partial\beta_1)^2+e^{-2\tau_2}(\partial\beta_2)^2\right)\\
 &+2e^{2D}\left(e^{-2\tau_1}z\partial_z\beta_1+e^{-2\tau_2}z\partial_z\beta_2\right)\biggr]\,,
 \end{split}
\end{equation}
where in the constant term we imposed the non-linear relation between $\sigma$ and $\tau$ given in \eqref{F3magVacuum2}. Moreover, as we did in previous examples, we absorbed a factor $e^{-2(\sigma+\sigma_1)+2D}$ in the kinetic terms in order to reproduce covariant expressions. We see that the present situation differs in a crucial way from the others we considered: because of the non-linear relation between $\sigma$ and $\tau$ in \eqref{F3magVacuum2}, factorizing out all the exponential scalings using the on-shell conditions is not automatic anymore.

\subsection{The off-shell action}\label{ref:offshellActionAdS3S3S3S1}

Let us compute the total off-shell action for $\mrm{AdS}_3\times S^3\times S^3\times S^1$ vacua. The space of variations is defined by the parameters $\sigma, \sigma_1, \tau_1, \tau_2,\tau_3, \Phi, \alpha, \beta_1, \beta_2$. As for the case of $\mrm{AdS}_3\times S^3\times \mrm{CY}_2$ vacua, the relevant terms contributing to the metric are the following
\begin{equation}
\begin{split}\label{IIBterms-offshell1}
 &S_{\text{IIB}}=S_{\text{EH}}+S_{\Phi}+S_{F_3}+S_{F_7}+\cdots\,,\\
  &S_{grav}=S_{\text{EH}}+S_{\Phi}=\frac{1}{2\kappa_{10}^2}\,\int d^{10}x\sqrt{-g}\,e^{-2\Phi}\left(R+4g^{mn}\partial_m\Phi\partial_n\Phi   \right)\,,\\
  &S_{flux}=-\frac{1}{2\kappa_{10}^2}\,\int d^{10}x\sqrt{-g}\,\left(\frac12\,|F_3|^2-\frac12|F_7|^2\right) \,.\\
  \end{split}
  \end{equation}
  We can obtain the explicit form of $S_{grav}$ from the general formula (\ref{dimRedGenera}), choosing the parameters $d=3$, $k_1=3$, $k_2=3$, $k_3=1$. We obtain the following three-dimensional action
  \begin{equation}
   \begin{split}\label{offshellActionMetricAdS3S3S3S1}
  S_{3,grav}&=\frac{1}{2}\int d^3x\,\,\sqrt{- g_{E}}\,\bigl(\tilde R_{E}+6\,e^{-2\tau_1+2D}+6\,e^{-2\tau_2+2D}-6e^{-2(\sigma+\sigma_1)+2D}z\partial_z\sigma_1\\
  &+2( \partial \,\sigma_f)^2-3\,( \partial\tau_1)^2
  -3\,( \partial\tau_2)^2-\,( \partial\tau_3)^2-( \partial D)^2
  \bigr)\,,
  \end{split}
\end{equation}
where we recall that $D=2\Phi-3\tau_1-3\tau_2-\tau_3$. Note that the variation of the external factor gives $( \partial \,\sigma_f)^2$ and not $( \partial \,\sigma)^2$. This is discussed in detail in sections \ref{sec:weylmoduli} and \ref{AdS3vacua}. The explicit expression for $\sigma_f$ is given in (\ref{onshellfinalppc}).

Including the contributions from flux variations \eqref{F3actionAdS3S3S3S1EF} and \eqref{F7actionAdS3S3S3S1EF2}, we obtain the total off-shell action:
\begin{equation}
\begin{split}\label{totalActionAdS3S3S3S1}
 S_{3}&=\,S_{3,grav}+S_{3,flux}=\\
 &=\frac12\int d^3x\,\,\sqrt{- g_{E}}\,\biggl[\tilde R_{E}+6\,e^{-2\tau_1+2D}+6\,e^{-2\tau_2+2D}-4\,e^{-2(\sigma+\sigma_1)+2D}\\
  &+2( \partial \,\sigma_f)^2-3\,( \partial\tau_1)^2
  -3\,( \partial\tau_2)^2-( \partial\tau_3)^2-( \partial D)^2-6e^{-2(\sigma+\sigma_1)+2D}z\partial_z\sigma_1\\
  &+2e^{-2(\sigma+\sigma_1)+2D}z\partial_z\alpha+2e^{2D}\left(e^{-2\tau_1}z\partial_z\beta_1+e^{-2\tau_2}z\partial_z\beta_2\right)\\
  &-\frac12(\partial\alpha)^2-\frac12e^{2(\sigma+\sigma_1)}\bigl(e^{-2\tau_1}(\partial\beta_1)^2+e^{-2\tau_2}(\partial\beta_2)^2\bigr)
 \biggr]\,.
 \end{split}
\end{equation}

We can now try to proceed by reducing this action requiring that the off-shell variations respect the relations \eqref{onshellAdS3S3S3S1}. We may recall them here expressed in terms of the lower-dimensional dilaton
\begin{equation}
\begin{split}\label{onshellAdS3S3S3S12}
 &\sigma+\sigma_1=\tau_1+\tau_2-\frac12\log\left(e^{2\tau_1}+e^{2\tau_2}\right)\,,\\
 &\alpha=2\sigma-\frac32\tau_1-\frac32\tau_2-\frac12\tau_3-\frac12D\,,\\
 &\beta_1=3\sigma+\sigma_1-\frac52\tau_1+\frac32\tau_2+\frac12\tau_3-\frac12D\,,\\
 &\beta_2=3\sigma+\sigma_1+\frac32\tau_1-\frac52\tau_2+\frac12\tau_3-\frac12D\,.
 \end{split}
\end{equation}
As we already noticed, $\tau_1, \tau_2$, are in a non-linear relation with the external volume variations $\sigma, \sigma_1$ (see the first relation in \eqref{onshellAdS3S3S3S12}). This issue can be observed at various levels. For instance, the exponential scalings in front of the derivative terms in $\beta_1, \beta_2$, do not drop out after imposing the on-shell conditions \eqref{onshellAdS3S3S3S12}, at least not naively. Alternatively, substituting the non-linear relation in the metric action leads to the same issue.
To deal with this problem, we can introduce a more appropriate parameterization of internal volume variations. Consider the following
\begin{equation}
\tau_+=\frac12(\tau_1+\tau_2)\;\;,\;\;\;\tau_-=\frac12(\tau_1-\tau_2) \;.
\end{equation}
In this notation, the non-linearity appears only in the $\tau_-$ variation. In fact, we can rewrite the on-shell conditions \eqref{onshellAdS3S3S3S12} as follows
\begin{equation}
 \begin{split}\label{onshellAdS3S3S3S1plusminus}
  &\sigma+\sigma_1=\tau_+-\frac12\log\big(2\cosh\left(2\tau_-\right)\big)\,,\\
  &\alpha=2\sigma-3\tau_+-\frac12\tau_3-\frac12D\,,\\
 &\beta_1=2\sigma+\frac12\tau_3-\frac12D-4\tau_--\frac12\log\big(2\cosh\left(2\tau_-\right)\big)\,,\\
 &\beta_2=2\sigma+\frac12\tau_3-\frac12D+4\tau_--\frac12\log\big(2\cosh\left(2\tau_-\right)\big)\,.
 \end{split}
\end{equation}
We can thus rewrite the off-shell action in this notation, obtaining
\begin{equation}
\begin{split}\label{totalActionAdS3S3S3S1plusminus}
 S_{3}&=\frac12\int d^3x\,\,\sqrt{- g_{E}}\,\biggl[\tilde R_{E}+6\,e^{-2(\tau_++\tau_-)+2D}+6\,e^{-2\tau_++2\tau_-+2D}-4\,e^{-2(\sigma+\sigma_1)+2D}\\
  &+2(\partial\sigma_f)^2-3( \partial\tau_++\partial\tau_-)^2
  -3\,( \partial\tau_+-\partial\tau_-)^2-( \partial\tau_3)^2-( \partial D)^2-6e^{-2(\sigma+\sigma_1)+2D}z\partial_z\sigma_1\\
  &+2e^{-2(\sigma+\sigma_1)+2D}z\partial_z\alpha+2e^{2D}\left(e^{-2(\tau_++\tau_-)}z\partial_z\beta_1+e^{-2\tau_++2\tau_-}z\partial_z\beta_2\right)\\
  &-\frac12(\partial\alpha)^2-\frac12e^{2(\sigma+\sigma_1)}\bigl(e^{-2(\tau_++\tau_-)}(\partial\beta_1)^2+e^{-2\tau_++2\tau_-}(\partial\beta_2)^2\bigr)
 \biggr]\,.
 \end{split}
\end{equation}

\subsection{The full metric over $\mrm{AdS}_3\times S^3\times S^3\times S^1$ vacua with $\tau_-=0$}\label{sec:AdS3S3S3S1taum}

In this section we compute the metric over variations for $\mrm{AdS}_3\times S^3\times S^3\times S^1$ vacua. More precisely, we compute the metric over infinite distance variations. The crucial point is that in the directions of $\tau_+$ and $\tau_-$, only variations of $\tau_+$ range to infinite distance. Variations of $\tau_-$ are bounded, and can only correspond to finite distance variations. To see this, note that holding $\tau_+$ fixed, and varying $\left|\tau_-\right|\rightarrow \infty$, we find either $\tau_1 \rightarrow -\infty$ or $\tau_2 \rightarrow -\infty$. Either of these regimes is not trustable within the supergravity approximation, since one of the spheres shrinks to zero size. We therefore have that the direction of $\tau_-$ cannot be taken non-compact, so it involves only finite size variations. In order to study infinite distance limits, we only need to consider $\tau_+$ variations, and so can set 
\be 
\tau_- = 0 \;.
\label{taum0}
\ee
We will maintain this condition throughout this section. 

With $\tau_-=0$, the volume variations of the two 3-spheres scale in the same way
 \begin{equation}
  ds^2_{10}=e^{2\sigma}d\hat s_3^2 +e^{2\tau_+}(d\hat s_{S_1^3}^2 +d\hat s^2_{S_2^3})+e^{2\tau_3}d\hat s^2_{S^1}\,.
 \end{equation}
This means that the non-linearity in the variations vanishes. The fundamental relations \eqref{onshellAdS3S3S3S1plusminus} take the following simplified form
 \begin{equation}
 \begin{split}\label{onshellAdS3S3S3S1plus}
  &\sigma+\sigma_1=\tau_+-\frac12\log2\,, \\
  &\alpha=2\sigma-3\tau_+-\frac12\tau_3-\frac12D\,,\\
  &\beta_1=\beta_2=2\sigma+\frac12\tau_3-\frac12D-\frac12\log2\,.
 \end{split}
\end{equation}
Moreover, we observe that the two contributions associated to magnetic fluxes define the same flux variation that we name as $\beta=\beta_1=\beta_2$.
The total off-shell action \eqref{totalActionAdS3S3S3S1plusminus} takes the following form
\begin{equation}
\begin{split}\label{totalActionAdS3S3S3S1plus}
S_{3}&=\frac12\int d^3x\,\,\sqrt{- g_{E}}\,\biggl[\tilde R_{E}+12\,e^{-2\tau_++2D}-4\,e^{-2(\sigma+\sigma_1)+2D}-6e^{-2(\sigma+\sigma_1)+2D}z\partial_z\sigma_1\\
  &+2(\partial\sigma_f)^2-6( \partial\tau_+)^2
  -( \partial\tau_3)^2-( \partial D)^2-\frac12(\partial\alpha)^2-\frac12(\partial\beta)^2\\
  &+2e^{-2(\sigma+\sigma_1)+2D}z\partial_z\alpha+2e^{-2(\sigma+\sigma_1)+2D}z\partial_z\beta
 \biggr]\,.
 \end{split}
\end{equation} The requirement that the linear derivative terms cancel takes the standard form
\begin{equation}\label{mixedtermsAdS3S3S3S1plus}
 \sigma_1=\frac{1}{3}(\alpha+\beta)\,.
\end{equation}
Specifically, solving together \eqref{mixedtermsAdS3S3S3S1plus} and \eqref{onshellAdS3S3S3S1plus} leads to the following on-shell relations
\begin{equation}
\begin{split}\label{finalvariationsAdS3S3S3S1plus}
&\tau_+=\frac{1}{6}(7\sigma-D+\log2)\,,\\
 &\alpha=\frac{1}{2}(-3\sigma-\tau_3-\log2)\,, \\
 &\beta=
 \frac{1}{2}(4\sigma+\tau_3-D-\log2) .\\
 \end{split}
 \end{equation}
 
 As in the case of $\mathrm{AdS}_3\times S^3\times \mathrm{CY}_2$ studied in section \ref{AdS3vacua}, we need to decompose the remaining variations into flux variations and a modulus variation. We proceed to calculate the metric over these two types of variations. 
 
\subsubsection{The metric over the moduli space}
\label{sec:AdS3S3S3S1metmodu}

The solution has a modulus, which we denote as $\chi$, that is a combination of $D$, $\tau_3$ and $\tau_+$. As in section \ref{AdS3vacua}, we should change basis to that of the modulus, and two orthogonal directions $\psi_1$ and $\psi_2$. A convenient choice of such basis is
\bea
\chi &=& \frac18 \left(D - 3\tau_3 +6 \tau_+ \right) \;, \nn \\
\psi_1 &=& \frac15 \left( 5 D + \tau_3 -2 \tau_+ \right) \;, \nn \\
\psi_2 &=&  \tau_3 + 3 \tau_+  \;.
\eea
With inverse relations
\bea
D &=& \frac12 \chi + \frac{15}{16} \psi_1 \;, \nn \\
\tau_3 &=& -\frac32 \chi + \frac{3}{16} \psi_1 +\frac{2}{5}\psi_2 \;, \nn \\
\tau_+ &=&  \frac12 \chi -\frac{1}{16} \psi _1 +\frac15 \psi_2  \;.
\eea
After this change of basis, the variation terms in (\ref{totalActionAdS3S3S3S1plus}) read
\be 
-6( \partial\tau_+)^2
  -( \partial\tau_3)^2-( \partial D)^2 = -4( \partial\chi)^2
  -\frac{15}{16}( \partial\psi_1)^2-\frac25 ( \partial \psi_2)^2  \;.
\ee 
We therefore obtain a metric over the moduli space of 
\be 
K_{\chi\chi} = 4 \;,
\ee 
matching known results, see for example \cite{Gukov:2004ym}. 

\subsubsection{The metric over AdS variations}
\label{sec:AdS3S3S3S1metadsvari}
 
To calculate the metric over the AdS variations, we need to start from the quadratic terms in the action (\ref{totalActionAdS3S3S3S1plus}), which read
\be 
  +2(\partial\sigma_f)^2-\frac12(\partial\alpha)^2-\frac12(\partial\beta)^2-\frac{15}{16}( \partial\psi_1)^2-\frac25 ( \partial \psi_2)^2 \;,
  \label{quadterms3s3s1d}
\ee 
where we dropped the kinetic term for the modulus $\chi$, which decouples. We then need to impose the on-shell conditions (\ref{finalvariationsAdS3S3S3S1plus}). These can be formulated in terms of the flux parameters, $\alpha$ and $\beta$, and the modulus $\chi$, as
\bea 
\sigma &=& \sigma_f + \frac12 \chi \;, \nn \\
\sigma_f &=& -\frac38 \alpha -\frac18 \log 4 \;, \nn \\
\psi_1 &=& -\frac{38}{15} \alpha - \frac{32}{15} \beta -\frac65 \log 4 \;, \nn \\
\psi_2 &=& -\alpha + \beta + \frac12 \log 2 \;.
\label{onshellfinalppc}
\eea 
Inserting the relations (\ref{onshellfinalppc}) into the terms (\ref{quadterms3s3s1d}), yields 
\be 
-\frac{637}{96} (\partial \alpha )^2 - \frac{28}{3} (\partial \alpha )(\partial \beta) -\frac{31}{6} (\partial \beta )^2 \;.
\ee 
The metric over the two-parameter space of AdS variations is therefore
\be 
K_{S^3\times S^1}=
\left(
\begin{array}{cc}
	\frac{637}{96} & \frac{14}{3}  \\
	\frac{14}{3} & \frac{31}{6} 
	\end{array}
\right) \;.
\ee 
As in the other cases, the metric is constant and positive definite, with eigenvalues
\be 
\lambda_1 \sim 10.6 \;,\;\; \lambda_2 \sim 1.2 \;.
\ee

\section{Summary}
\label{sec:dis}

In this work we studied the notion of a metric over families of AdS vacua in string theory. The procedure for calculating the metric follows the ideas introduced in \cite{Li:2023gtt,Palti:2024voy}, but is developed further to account for increasingly complicated families of solutions. In particular, we studied families of solutions which are multi-parameter. This means that the metric is not over a one-dimensional space, and so can have additional features. Further, it allows for moduli directions on top of directions varying the AdS radius. 

Perhaps the most important feature of the metric is its positivity. Like the solutions studied in \cite{Li:2023gtt,Palti:2024voy}, we find that all the solutions studied here also have a positive definite metric. The metric therefore yields a well-defined distance along any path in the space of solutions. 

We find that the metric is also flat, something which is not trivial in this multi-dimensional case. More precisely, we find that the metric over the space of solutions which come in infinite families is flat. So the metric measuring distances which can become infinite. When there is a solution parameter which can only vary over a finite range, the metric is not flat. We encounter such an example in the case of $\mrm{AdS}_3\times S^3\times S^3\times S^1$ solutions. In fact, we have not developed a procedure for calculating the metric including such finite, or compact, directions in solution space. Instead, we truncated these compact directions, keeping only those which come in infinite families. It would be interesting to understand how to calculate a metric over all the parameter space, including the compact directions. 

Moduli are directions in the space of solutions that do not vary the AdS radius, so  which have no potential associated to them. We encountered such directions in the cases of $\mrm{AdS}_3\times S^3\times \mathrm{CY}_2$ and $\mrm{AdS}_3\times S^3\times S^3\times S^1$. We find that the AdS variations and the moduli variations factorise, and are naturally treated separately. The moduli space metric is recovered as usual, since our procedure for calculating the metric is identical to the procedure for calculating the moduli space metric in the case when the variations are moduli. The procedure for calculating the metric over AdS variations needs to be slightly modified to account for the presence of moduli. It requires a certain ordering of the Weyl rescaling and the metric variations, which arises only when moduli are present. The resulting metric is, as in other cases, flat and positive. 

Our results lend further evidence to the conjecture that the metric over families of AdS solutions, as calculated through the introduced prescription, is always positive \cite{Palti:2024voy}. In fact, we expect that this can eventually be proven. More generally, we have found that the procedure introduced in \cite{Li:2023gtt} for calculating a metric over the space of AdS solutions has held up to tests very well. We summarise in table \ref{tab:solutions} the families of solutions studied so far, in this paper and in \cite{Li:2023gtt,Palti:2024voy}. In all the cases the prescription yields a well-defined and precise notion of distance between solutions. We expect that this notion of distance is indeed physically meaningful, and that the AdS distance conjecture \cite{Lust:2019zwm} is capturing the behaviour of string theory with respect to this distance. 

\begin{table}[h!]
\def\arraystretch{1.5}%
\begin{center}
	\begin{tabular}{|c|c|c|}
		\hline
		Solutions type & Properties & Calculated Metrics \\
		\hline
		\hline
		$\mathrm{AdS}_4 \times S^7$ (M-theory) & 1-parameter & $K=\frac{1563}{8}$\\
		\hline
		$\mathrm{AdS}_7 \times S^4$ (M-theory)& 1-parameter& $K=\frac{516}{5}$\\
		\hline
		$\mathrm{AdS}_4 \times \mathrm{CY}_3 $ (IIA)$^*$ & \begin{tabular}{c} 1-parameter \\ Additional fluxes \\ Scale-separated  \end{tabular} & $K=\frac{3376}{27}$\\
		\hline
		$\mathrm{AdS}_4\times \mathbb{CP}^3$ (IIA) & \begin{tabular}{c} 2-parameter  \end{tabular} & 
		\begin{tabular}{c}
		$K=\begin{pmatrix}
2196 & -546 \\
-546 & 136
\end{pmatrix}$ 
\\
$\lambda_1 \sim 2332 \;,\;\; \lambda_2 \sim 0.2$
\end{tabular}
\\
		\hline
		$\mrm{AdS}_3\times S^3\times \mathrm{CY}_2$ (IIB)& \begin{tabular}{c} 2-parameter \\ Modulus  \end{tabular} & \begin{tabular}{c} 
		$K=\left(
\begin{array}{cc}
	\frac{113}{24} & \frac{89}{24}  \\
	\frac{89}{24} & \frac{113}{24} 
	\end{array}
\right)$ 
\\
$\lambda_1 = \frac{101}{12}\;,\;\; \lambda_2 = 1\;.$ 
\\
$K_{\chi\chi} = 4\;.$
\end{tabular}
\\
		\hline
		$\mrm{AdS}_3\times S^3\times S^3\times S^1$ (IIB)& \begin{tabular}{c} 2-parameter \\ Modulus \\ Compact parameter \end{tabular} & 
		\begin{tabular}{c}
		$K=\left(
\begin{array}{cc}
	\frac{637}{96} & \frac{14}{3}  \\
	\frac{14}{3} & \frac{31}{6} 
	\end{array}
\right) $
\\
$\lambda_1 \sim 10.6 \;,\;\; \lambda_2 \sim 1.2\;.$
\\
$K_{\chi\chi} = 4\;.$
\end{tabular}
\\
		\hline
	\end{tabular}
	\caption{Table showing different families of AdS solutions for which the metric has been calculated, in this paper and in \cite{Li:2023gtt,Palti:2024voy}. For each family we describe some of its key properties, showing that metrics have been calculated for increasingly complex settings. The metrics are given in the final column and denoted as $K$, and its eignevalues are denoted by $\lambda_1$ and $\lambda_2$. A moduli space metric is denoted by $K_{\chi\chi}$. Note that the metrics depend on the coordinates chosen, so which parameters are used to parameterise the space of solutions. However, their positivity is independent of this choice, and they yield a well-defined distance for any parameterisation of the solutions.
	\newline
	$^*$ The $\mathrm{AdS}_4 \times \mathrm{CY}_3$ case is not a full ten-dimensional solution, but defined through the effective four-dimensional action only \cite{DeWolfe:2005uu}.}
	\label{tab:solutions}
	\end{center}
\end{table}

\vskip 20pt
	\noindent {\bf Acknowledgements:} We thank Niall Macpherson for useful discussions. The work of EP is supported by the German Research Foundation (DFG) through a German-Israeli Project Cooperation (DIP) grant ``Holography and the Swampland", and by the Israel planning and budgeting committee grant for supporting theoretical high energy physics.
The work of NP is supported by INFN through the project ``Understanding gravity via gauge theories, supergravity and strings''.

\newpage
\appendix

\section{$\mathbb{CP}^3$ space and $F_2$ flux actions}
\label{app:F2fluxes}

In this appendix we derive the on-shell action for $F_2$ flux and the corresponding off-shell action for flux variations. Let us recall some important properties of $\mathbb{CP}^3$. This is a K\"ahler Einstein manifold with an $\text{SO}(6)$ invariant form $J$. On $\mathbb{CP}^3$ we have the following invariant forms: $J$, $J\wedge J$ and $J\wedge J \wedge J$. Specifically, we have the fundamental relations
\begin{equation}\label{fundamentalJrel}
 J\wedge J=2\,\hat\star_6J\,,\qquad \qquad \,\hat\star_6\left(J\wedge J \right)=2J\,,
\end{equation}
where we use the notation $\hat \star_6$ for the Hodge dual operator on the unit radius $\mathbb{CP}^3$ with metric $d\hat s^2_6=e^{-2\tau}\,ds^2_6$ (we use the notation of \eqref{AdS4CP4metricgeneral}).
Using the above equations one can obtain the standard relation for the volume form of unit radius $\mathbb{CP}^3$,
\begin{equation}\label{volumeCP3}
 J\wedge J\wedge J=6\,\text{vol}_{\mathbb{CP}^3}\,.
\end{equation}
The equations \eqref{fundamentalJrel} can be used to relate the on-shell $F_2$ flux in \eqref{AdS4CP3} to its electric partner $F_8$. We remind that, for the moment, all the physical scaling are treated as constant parameters. We can start from the expression for $F_8$ in \eqref{AdS4CP4fluxgeneral},
\begin{equation}
\begin{split}\label{app:F2F8fluxes}
&F_8=dC_7=-3e^{\beta}\text{vol}_4\wedge J\wedge J\;,\\
&F_2=\star_{10} F_8=-3e^\beta\,\star_4\text{vol}_4\wedge \star_6\left (J\wedge J\right)=6\,e^{\beta-4\sigma-\sigma_1-2\tau}\,J\,,
\end{split}
\end{equation}
where in the above expressions we used that $\star_4 \text{vol}_4=-e^{-4\sigma-\sigma_1}$ and $\star_6=e^{-2\tau}\hat {\star}_6$. Finally, to obtain $F_2$ we applied the second relation in \eqref{fundamentalJrel}.

We can compute the contribution to the vacuum energy of the $F_2$ flux, describing the last contribution in \eqref{AdS4CP3action}. To this aim, we must evaluate the $F_2$ on-shell action over \eqref{app:F2F8fluxes}, obtaining
\begin{equation}\label{app:F2action1}
 -\frac{1}{4\kappa^2_{10}}\int d^{10}x\sqrt{-g}\,|F_2|^2=-\frac{9}{ \kappa^2_{10}}\,\int d^4x\,\sqrt{-g_4}e^{6\tau+2\beta-8\sigma-2\sigma_1-8\tau}\,\int J\wedge \hat\star_6J\,,
\end{equation}
where we wrote $\int d^6y\,\sqrt{\hat g_6}\,J_{ab}\,J_{cd}\,\hat g_6^{ac}\,\hat g_6^{bd}=\int J\wedge \hat \star_6 J$ with $\{y^a\}$ coordinates over $\mathbb{CP}^3$. Now we can use the first equation in \eqref{fundamentalJrel} and the expression \eqref{volumeCP3} for the $\mathbb{CP}^3$ volume form, to obtain the following result
\begin{equation}\label{app:F2action2}
 =-\frac12\,\int d^4x\sqrt{-g_4}e^{6\tau+2\beta-8\sigma-2\sigma_1-8\tau}\,54\,,
\end{equation}
where we fixed $\kappa_{10}^2=\text{Vol}(\mathbb{CP}^3)=\int  d^6y\,\sqrt{\hat g_6}$. Finally, we can go to the Einstein frame $ds^4_{4}=e^{2D}ds^2_E$ with $D=\Phi-3\tau$.
We thus obtain the contribution of the 2-flux to the 4d effective action \eqref{AdS4CP3action}
\be 
-\frac12\,\int d^4x\sqrt{-g_E}e^{2\beta-8\sigma-2\sigma_1-8\tau+2\Phi+2D}\,54 \;.
\ee

We can now derive the off-shell action of variations of the electric flux $F_8$, introduced in \eqref{C7andF8}. In fact, as we argue in section \ref{Sec:fluxes}, for magnetic fluxes, off-shell variations must be taken over the electric dual flux, in our case $F_8$. In particular, equation \eqref{idHodge} implies that the contribution to the off-shell action of a dual flux is given by the standard flux action with an overall sign switched
\be
 -\frac{1}{4\kappa^2_{10}}\int F_2\wedge \star_{10}F_2=\frac{1}{4\kappa^2_{10}}\int F_8\wedge \star_{10}F_8 \;.
 \ee 

We may recall here our definition of the gauge potential $C_7$ and the corresponding flux variation,
\begin{equation}
\begin{split}\label{app:C7andF8}
& C_7=-e^\beta z^{-3}\,dx^0\wedge dx^1 \wedge dx^{2}\wedge J\wedge J\,,\\
&F_8=dC_7=-\,e^{\beta}\left(3-z\partial_z\beta  \right)\text{vol}_4\wedge J\wedge J\,,
\end{split}
\end{equation}
where now all the scaling parameters are taken as local functions along the $z$ direction. We can compute the off-shell action for $F_8$ variations,
\begin{equation}
\begin{split}\label{app:F8action1}
 \frac{1}{4\kappa^2_{10}}&\int d^{10}x\sqrt{-g}\,|F_8|^2=\frac{1}{4\kappa^2_{10}}\,\int e^{2\beta}(3-z\partial_z \beta)^2\,(\text{vol}_4\wedge J^2)\wedge (\star_4 \vol_4\wedge\star_6 J^2)\\
&=-\frac{1}{4\kappa^2_{10}}\int d^4x\sqrt{-g_4}\,e^{2\beta-8\sigma-2\sigma_1}(3-z\partial_z \beta)^2J^2\wedge \star_6 J^2  \;,
 \end{split}
\end{equation}
where we wrote the four-dimensional part of the integral using the relation $\star_4 \text{vol}_4=-e^{-4\sigma-\sigma_1}$. The internal six-dimensional part of the integral can be treated as follows. We can use the second fundamental relation \eqref{fundamentalJrel}, together with $\star_6=e^{-2\tau}\hat {\star}_6$ and the formula for the volume form \eqref{volumeCP3}. In this way we obtain the following identities $\int J^2\wedge \star_6 J^2=\int e^{-2\tau}J^2\wedge \hat\star_6 J^2=\int 2e^{-2\tau}J\wedge J^2=\int 12e^{-2\tau}\text{vol}_{\mathbb{CP}^3}=\int12e^{-2\tau}\sqrt{\hat g_6}\,d^6y$. We can then rewrite the action as 
\begin{equation}\label{app:F8action2}
 =-\frac{1}{2}\int d^4x\sqrt{-g_4}\,e^{2\beta-8\sigma-2\sigma_1-2\tau}\;6\left(3-z\partial_z \beta\right)^2\,,
\end{equation}
which is precisely the relation written in \eqref{F2action10D} with $\kappa_{10}^2=\text{Vol}(\mathbb{CP}^3)=\int  d^6y\,\sqrt{\hat g_6}$.
Finally we can express the result in the Einstein frame $ds^4_{E}=e^{-2D}ds^2_4$ with $D=\Phi-3\tau$. We thus obtain the following result
\begin{equation}
 S_{4, F_8}=-\frac{1}{2}\int d^4x\sqrt{-g_E}\,e^{2\beta-8\sigma-2\sigma_1-8\tau+2\Phi+2D}\;6\left(3-z\partial_z \beta\right)^2\,,
 \end{equation}
which is the action written in  written in \eqref{F2action}.

	\bibliographystyle{jhep}
	\bibliography{metricdistance}
\end{document}